\documentclass[aps,pra,onecolumn,superscriptaddress,preprintnumbers,secnumroman,nofootinbib,12pt]{revtex4-2}
\usepackage[T1]{fontenc}
\usepackage{epsfig}
\usepackage{graphicx}
\usepackage[english]{babel}
\usepackage{hyphenat}
\usepackage{amsmath}
\usepackage{amssymb}
\usepackage{bbold}
\usepackage{mathtools}
\usepackage{mathrsfs}
\usepackage{slashed}
\usepackage{epstopdf}
\usepackage{xcolor}
\usepackage{braket}
\usepackage{booktabs}
\definecolor{lcolor}{rgb}{0.,0.0,0.}
\definecolor{citcolor}{rgb}{0,0.,0.5}
\usepackage[breaklinks,colorlinks,urlcolor=blue,citecolor=blue,linkcolor=blue]{hyperref}
\usepackage{multirow}
\usepackage{ltablex}
\usepackage[normalem]{ulem} 
\usepackage{changepage} 
\usepackage{lipsum}

\usepackage[a4paper,margin=1in]{geometry}
\usepackage{setspace}
\usepackage[e]{esvect}

\usepackage{media9}

\def\cP{{\cal P}}

\def\cN{{\cal N}}

\def\cO{{\cal O}}

\newcommand{\secn}[1]{Section~1}
\newcommand{\appn}[1]{Appendix~1}

\long\def\comment#1{ }

\def\and{\quad\text{and}\quad}

\def\q{{\boldsymbol q}}
\def\0{{\boldsymbol 0}}
\def\1{{\boldsymbol 1}}
\def\p{{\boldsymbol p}}

\def\b{{\boldsymbol b}}
\def\0{{\boldsymbol 0}}

\def\bn{{\boldsymbol n}}

\newcommand{\cF}{\mathcal{F}}

\renewcommand\b{\beta}
\renewcommand\d{\delta}

\renewcommand\o{\omega}

\newcommand\m{\mu}

\renewcommand{\part}{{\rm part}}

\newcommand{\be}{\begin{equation}}
\newcommand{\ee}{\end{equation}}
\newcommand{\bes}{\begin{subequations}}
\newcommand{\ees}{\end{subequations}}
\newcommand{\bea}{\begin{eqnarray}}
\newcommand{\eea}{\end{eqnarray}}

\newcommand{\pa}{\partial}

\newcommand{\nn}{\nonumber \\}

\def\bea#1\eea{\begin{align}#1\end{align}}
\newcommand{\bef}{\begin{figure}[h!tb]\centering}
\newcommand{\eef}{\end{figure}}

\allowdisplaybreaks

\newcommand\n{\mathbf n}

\begin{document}
\newgeometry{
  top=.5in,bottom=.9in,
  left=.5in,right=.5in,
  hcentering,
  includeheadfoot
}

\title{\LARGE Hydrodynamics and Energy Correlators}


\author{Jo\~{a}o Barata}
\email{joao.lourenco.henriques.barata@cern.ch}
\affiliation{European Organization for Nuclear Research (CERN),  Theoretical Physics Department, CH-1211 Geneva, Switzerland}

\author{Matvey V. Kuzmin}
\email{matvei.kuzmin@physics.ox.ac.uk}
\affiliation{Rudolf Peierls Centre for Theoretical Physics, Department of Physics,
Parks Road, Oxford, OX1 3PU, UK}

\author{Ian Moult}
\email{ian.moult@yale.edu}
\affiliation{Department of Physics, Yale University, New Haven, CT 06511}

\author{Andrey V. Sadofyev}
\email{andrey.sadofyev@ehu.eus}
\affiliation{Department of Physics, University of the Basque Country UPV/EHU, P.O. Box 644, 48080 Bilbao, Spain}
\affiliation{IKERBASQUE, Basque Foundation for Science, Plaza Euskadi 5, 48009 Bilbao, Spain}
\affiliation{Laboratório de Instrumentação e Física Experimental de Partículas (LIP), Av. Prof. Gama Pinto, 2, 1649-003 Lisbon, Portugal}

\author{Jo\~{a}o M. Silva}
\email{joao.m.da.silva@tecnico.ulisboa.pt}
\affiliation{Laboratório de Instrumentação e Física Experimental de Partículas (LIP), Av. Prof. Gama Pinto, 2, 1649-003 Lisbon, Portugal}
\affiliation{Departamento de Física, Instituto Superior Técnico (IST), Universidade de Lisboa, Av. Rovisco Pais 1, 1049-001 Lisbon, Portugal}
\affiliation{Departamento de Física Teórica y del Cosmos, Universidad de Granada, Campus de Fuentenueva, E-18071 Granada, Spain}
\preprint{CERN-TH-2026-094}

\begin{abstract}
We study energy-energy correlators (EECs) in many-body quantum states, focusing on the matter produced in the aftermath of heavy-ion collisions. 
We analyze the angular structure of EECs in the collinear limit and identify a sequence of dynamical regimes. At the largest angular separations within the small-angle regime, the observable is dominated by disconnected contributions, leading to a classical scaling determined by the collective flow of the medium. We explicitly construct this contribution
for hadrons produced from a hydrodynamic medium described by boost-invariant Gubser flow, obtaining the angular dependence of the EEC analytically. We further consider azimuthal perturbations to this flow, illustrating how EECs can be used to probe anisotropies in the initial state. At smaller angular separations, connected contributions become increasingly important. We argue that in this regime the EEC is controlled by collective hydrodynamic modes. 
The resulting angular behavior is similar to the one identified in the EECs of heavy and large-charge states of conformal field theories. At even smaller angles, this regime is expected to match onto the structure determined by the light-ray operator product expansion, before eventually crossing over to the smallest-angle behavior characteristic of dilute hadronic matter. Altogether, these results provide a unified picture of the angular structure of EECs in many-body QCD states and suggest new observables sensitive to the properties of matter in heavy-ion collisions.
\end{abstract}

\maketitle

\clearpage
\restoregeometry
\tableofcontents

\clearpage

\section{Introduction}
\label{sec:introduction}

Understanding how collective many-body behavior emerges from microscopic dynamics remains one of the central problems in quantum field theory (QFT). This question spans a wide range of physical systems, from strongly coupled condensed-matter systems to astrophysical environments, and lies at the heart of the broader effort to understand how complex macroscopic matter emerges from fundamental degrees of freedom. In high-energy physics, it has been studied most prominently in the context of gauge theories, which form the theoretical backbone of the Standard Model, with particular emphasis on Quantum Chromodynamics (QCD). Although QCD governs the microscopic dynamics of the partonic constituents of ordinary nuclear matter, explaining the emergence of collective behavior from this description remains a major challenge. 

Remarkably, this question can be experimentally studied through deep inelastic scattering and heavy-ion collisions (HIC), which provide access to the structure and dynamics of QCD matter across a wide range of scales. For example, in HICs, this includes the hot and dense matter created in conditions similar to those of the early Universe~\cite{Busza:2018rrf,Arslandok:2023utm}. A major outcome of the HIC program has been the discovery that the quark-gluon plasma (QGP) created in such collisions behaves as an almost ideal fluid, with relativistic hydrodynamics providing a successful effective description of much of its bulk evolution, see e.g.~\cite{Shen:2020mgh,Heinz:2024jwu}. This collective behavior has been established most clearly through the observed anisotropic flow of the final-state hadrons.

Despite this progress, a complete quantitative understanding of the real-time dynamics of the bulk matter produced in these experiments remains out of reach. While hydrodynamic observables provide strong evidence for the emergence of a locally equilibrated fluid, they are less sensitive to the out-of-equilibrium dynamics preceding the QGP phase. As a result, current experimental data do not yet decisively discriminate between competing scenarios of early-time thermalization and hydrodynamization~\cite{Berges:2020fwq}. These limitations motivate the search for new theoretical tools and observables capable of probing the properties and structure of the many-body states formed throughout the evolution of QCD matter in HICs, thereby enabling a more direct connection between microscopic descriptions and experimentally accessible measurements across different dynamical regimes.

In this context, asymptotic energy-flow operators 
\begin{align}\label{eq:def_Ec}
\mathcal{E}(\mathbf{n}) = \lim_{r\to\infty} r^2 \int_0^\infty dt\, n^i T^{0i}(t,r\mathbf{n})\, ,
\end{align}
and the energy correlators (ECs) constructed from them  have attracted significant attention in recent literature, following the illustration that they can be used as practical jet substructure observables \cite{Dixon:2019uzg,Chen:2020vvp,Komiske:2022enw}, see \cite{Moult:2025nhu} and references therein for a review. These non-local operators measure the energy flux through a distant detector oriented along the direction $\mathbf n$, and provide the basic building blocks for a class of event shapes and correlation observables in collider physics. Originally introduced in QCD~\cite{Basham:1978bw,Basham:1979gh,Belitsky:2001ij} as a natural way to formulate infrared-safe measurements, they have more recently also become central in formal studies of QFT, where expectation values of products of such operators can be analyzed using operator product expansion (OPE) based methods~\cite{Hofman:2008ar}. This dual role makes correlators of energy-flow operators particularly powerful: they are closely tied to experimentally accessible energy distributions, while at the same time admitting a transparent formulation within QFT.

The application of energy-flow operators in QCD has seen numerous  developments in precision studies of the microscopic structure of the theory, see e.g. \cite{Lee:2022uwt,Chen:2019bpb,Craft:2022kdo,Holguin:2022epo,Devereaux:2023vjz,Chen:2022swd,Lee:2023npz,Chicherin:2024ifn,Chen:2023zzh,Kang:2023big,Gao:2023ivm,Chen:2024bpj,Chen:2024nfl,Chen:2024nyc,Budhraja:2024tev,Liu:2024lxy,Alipour-fard:2024szj,Barata:2024wsu,Mantysaari:2025mht,Alipour-fard:2025dvp,Chen:2025rjc,Guo:2025zwb,Guo:2025qnz,Gao:2025cwy,Kang:2026hig,Budhraja:2026pyi,Riembau:2025wjc,  Riembau:2025isw,Fu:2024pic,Gao:2026xuq,Holguin:2026vld,Jaarsma:2025tck,Budhraja:2026pyi,Chang:2025kgq,Jaarsma:2023ell,Chen:2023zlx,Chen:2025rjc,Electron-PositronAlliance:2025fhk,Chen:2026hmd} for recent discussions. Their extension to genuinely many-body states, however, is much more recent and has so far emerged across several largely separate lines of investigation.
In CFTs, existing studies have focused on heavy~\cite{Chicherin:2023gxt} or large-charge~\cite{Cuomo:2025pjp} states, where correlation functions can be organized systematically in an expansion controlled by the parameters of the state. These studies have provided a map of the angular structure of many-body ECs and their relation to underlying properties of the theory. Another line of investigation relies on applying the Eigenstate Thermalization Hypothesis (ETH)~\cite{Deutsch:1991msp,Srednicki:1994mfb,Deutsch:2018ulr} to locally equilibrated \textit{fireball} states~\cite{Delacretaz:2018cfk}, allowing one to constrain fluctuations around equilibrium through general positivity bounds. From the QCD phenomenological perspective, energy correlator observables were first proposed to study jets in HICs in \cite{Andres:2022ovj}.\footnote{See~\cite{Krasnitz:1998ns} for an earlier proposal to use energy correlators to study many-body states in HICs.} 
From this perspective, energy correlator observables have so far focused primarily on its coupling to jet-like excitations produced in the same event, see~\cite{Andres:2022ovj,Andres:2023xwr,
Andres:2024ksi,Yang:2023dwc,Barata:2023bhh,Singh:2024vwb,Xing:2024yrb,Bossi:2024qho,Andres:2024xvk,Barata:2025uxp,Apolinario:2025vtx,Liu:2025ufp,Ke:2025ibt,Budhraja:2025ulx,Apolinario:2026hff,Barata:2025zku,Kudinoor:2026wcs,Krasnitz:1998ns} for jet-quenching studies using energy-flow operators. 
The first dedicated studies of hydrodynamic response for the energy correlators were performed in~\cite{Bossi:2024qho} and~\cite{Yang:2023dwc}, using the Hybrid model~\cite{Casalderrey-Solana:2016jvj,Casalderrey-Solana:2014bpa} and LBT~\cite{He:2015pra,Cao:2016gvr} frameworks. The analytic investigation of hydrodynamics effects on energy correlators in HICs was initiated in~\cite{Barata:2024ukm,Barata:2025fzd} where it was shown that the collective flow of the medium induced by a jet contributes to the measured energy correlators already at the classical level, resulting in a characteristic angular scaling, in agreement with the results of dedicated Monte Carlo simulations~\cite{Bossi:2024qho,Yang:2023dwc}, see also~\cite{Zhao:2025ogc} and \cite{Duan:2026icj}. While this disconnected contribution to ECs was first identified in this context, the underlying geometric correlation is more general and does not rely on the presence of jets. Taken together, these developments indicate that energy correlators provide a promising probe of many-body dynamics, while a systematic study of bulk QCD matter in HICs using these observables is still missing.

In this work, we initiate a systematic study of bulk QCD matter produced in HICs using energy-flow operators. Our analysis focuses on the two-point function of energy-flow operators, namely the energy-energy correlator (EEC), although the discussion can be straightforwardly generalized to higher-point functions. In this way, we connect the HIC setting to the broader discussion of EECs on many-body states. We argue that, when measured on a nearly equilibrated state, the EEC exhibits a characteristic angular structure spanning several physical regimes. At the largest angular separations within the small-angle regime, the observable is dominated by a disconnected contribution associated with uncorrelated classical energy flow. For a near-equilibrium medium, it is captured by hydrodynamics, leading to a characteristic linear scaling with the opening angle between the detectors. We provide an explicit analysis of this behavior for Gubser flow~\cite{Gubser:2010ui,Gubser:2010ze}, which offers an analytic model of an expanding, transversely finite QGP state produced in HICs in Sec.~\ref{sec:Gubser_flow_sec}. At intermediate angles, connected contributions can become significant, leading to a different angular behavior governed by the collective modes of the near-equilibrium medium, see Sec.~\ref{sec:ETH}. At smaller angles, the scaling becomes increasingly sensitive to the microscopic properties of the theory and should be matched onto the celestial OPE, at least if the theory becomes approximately conformal in that region. Finally, if dimensional transmutation is present, as in QCD, and the associated scale is well separated from the characteristic scales of the collective thermal sector, the smallest angular separations probe the late-time hadronic regime, in which the EEC exhibits the linear rise characteristic of a dilute hadronic gas. We stress, however, that unlike in idealized conformal setups, the information encoded in all of these angular regimes is ultimately transmitted to the detector by hadronic final states. Hadronization therefore enters the experimental realization of the full angular structure, even though only the smallest-angle regime may be directly governed by dilute hadronic physics.

\begin{figure}[h!]
    \centering
    \includegraphics[width=0.9\linewidth]{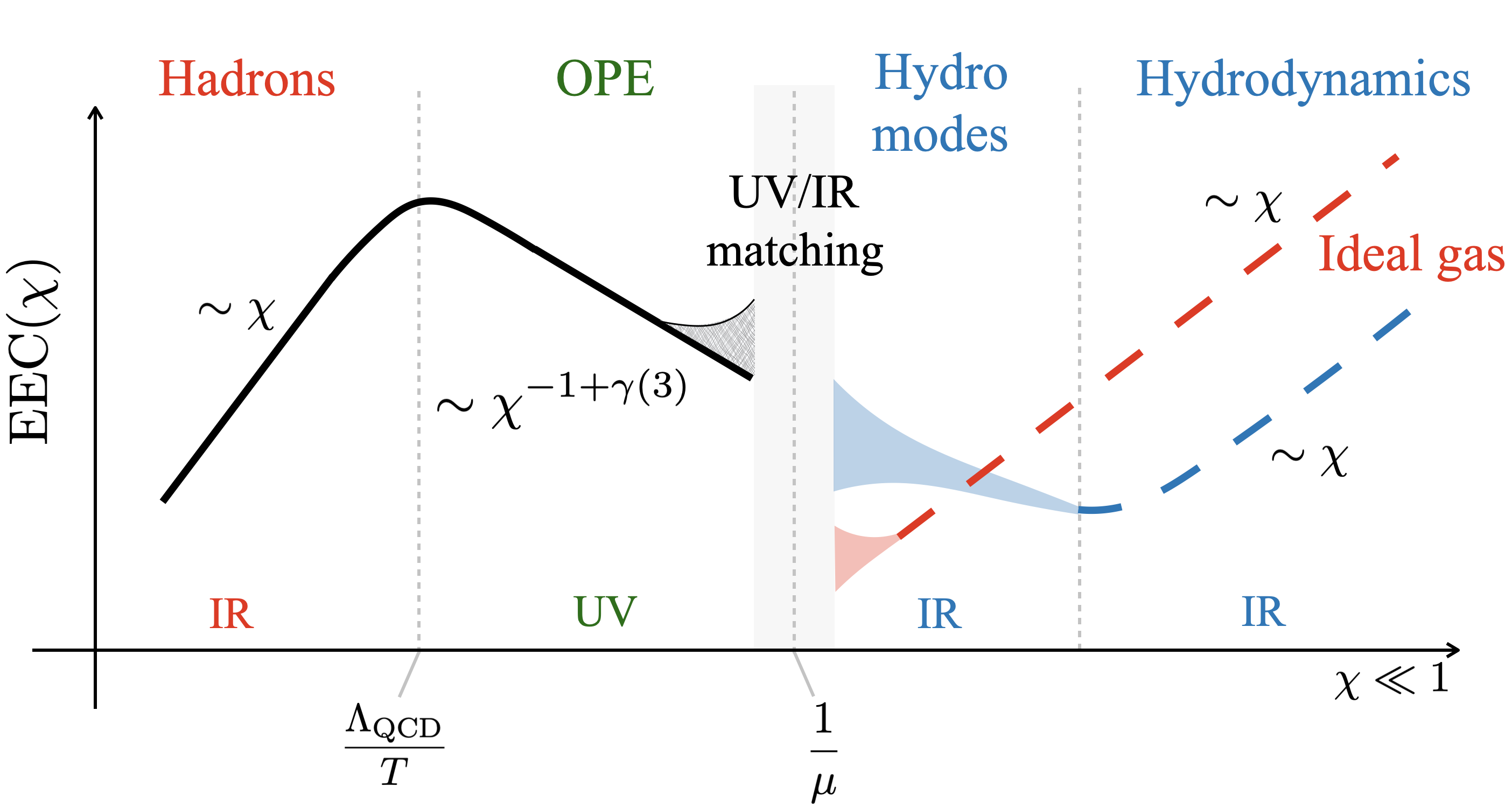}
    \caption{Schematic angular structure of the EEC measured on a bulk matter state for angular separations between detectors $\chi\ll 1$. Dashed vertical lines separate the physical regimes discussed in the text. At the smallest angles, the EEC exhibits the linear rise characteristic of a dilute hadronic gas, reflecting its effectively ballistic nature. At larger angles, the EEC is expected to be described by the leading OPE terms, with higher-twist contributions becoming relevant as the angle grows, as indicated by the hatched region. Closer to $\mu^{-1}$, the OPE must be matched onto an effective description of the medium, with the corresponding UV/IR matching region indicated by the shaded band. Here $\mu\sim Tr_{\rm ch}$, where $r_{\rm ch}$ is the characteristic size of the system and $T$ is its characteristic temperature. At even larger angles, the EEC becomes sensitive to collective modes of the medium and eventually enters the hydrodynamic regime, where it is again dominated by uncorrelated classical energy flow. The dashed red line illustrates the characteristic behavior expected for a free-streaming gas, where the linear scaling extends through the collective region.}
    \label{fig:cartoon}
\end{figure}

Figure~\ref{fig:cartoon} schematically summarizes the angular structure of the EEC discussed above; further details are given in Sec.~\ref{sec:ETH}. Several general lessons are worth emphasizing:
\begin{itemize}
\item
The angular structure of the correlator reflects an IR/UV/IR sequence governed by confinement dynamics, the celestial OPE, and collective dynamics of the medium. The departure from leading twist scaling in the OPE due to enhanced higher twist effects from the presence of a nuclear medium was recently highlighted in~\cite{Andres:2024xvk}, see also \cite{Devereaux:2023vjz,Yang:2023dwc,Fu:2024pic,Barata:2023bhh}. The full matching between the OPE regime and the collective infrared dynamics demands a dedicated analysis, which is beyond the scope of the present work and is indicated by the gray band in Fig.~\ref{fig:cartoon}. Such a matching would require relating the relevant operator content and scaling behavior on both sides of the transition, along the lines discussed in~\cite{Chang:2025kgq,Chang:2025zib}.

\item
The large-angle behavior discussed here also applies to jet-based EEC measurements in HICs, as previously analyzed in~\cite{Barata:2024ukm,Barata:2025fzd}. More generally, this highlights that the HIC setting naturally extends the usual celestial OPE picture~\cite{Andres:2024xvk} by incorporating the large-angle behavior associated with many-body dynamics. When a clear separation of scales exists, as schematically illustrated in Fig.~\ref{fig:cartoon}, the OPE need not converge at all angles and must be supplemented by the collective response of the medium.  While the observable remains finite and the operator-level discussion of the celestial OPE remains meaningful, applying it to a matter-plus-jet state is highly nontrivial. Nevertheless, in such a situation, we would expect that the departure from leading twist scaling should be a robust signature. 
It will be extremely interesting to have exactly solvable models of the transition between the OPE regime and the hydro regime. 

In realistic situations, the separation between the scales indicated in the figure and those associated with the jet may not be sharp. As a result, the coefficients appearing in the angular expansion are expected to receive contributions both from the perturbative jet cascade and from the collective bulk flow, which in principle could be disentangled through their different parametric dependencies, see the discussion in~\cite{Barata:2025fzd}. 
More concretely, at the classical level, the jet based EEC in HICs has the structure
\begin{align}~\label{eq:ooo}
    {\rm EEC} = \frac{a_2}{\chi} &+ a_4 \chi + a_6 \chi^3 + \mathcal{O}(\chi^5) \nn 
    &+  b_4 \chi +b_6 \chi^3 + \mathcal{O}(\chi^5) \, .
\end{align}
The terms on the first line describe the jet fragmentation and can be treated within a celestial OPE based discussion \cite{Andres:2024xvk}, i.e. twist expansion. The terms on the second line are controlled by hydrodynamics, and thus require a different treatment, which we follow to detail. The degenerate form of Eq.~\eqref{eq:ooo} prevents an immediate separation of the $\{a\}$ and $\{b\}$ coefficients. This motivates the theoretical exploration of both contributions, to understand how they can be experimentally disentangled. For initial investigations, see \cite{Barata:2025fzd}.

\item
Several strategies may help isolate the different regimes and use EECs as direct probes of bulk matter. On the theory side, contributions arising from distinct physical mechanisms are expected to exhibit different parametric scalings with the properties of the medium, which can in principle be estimated and used to organize the different contributions to the EEC, see also the discussion in \cite{Barata:2025fzd}. Such dependences could be inferred from experimental data, enabling more detailed tests of the predicted angular behavior. More broadly, this opens the possibility of using EECs to access soft many-body dynamics in HICs in a way complementary to more traditional observables. This program could be further expanded by a systematic scan across collision systems of different sizes, ranging from $PbPb$ and $AuAu$ to $OO$, $pA$, and $pp$, where the relative importance of the different dynamical regimes is expected to vary significantly. Such a scan could also help disentangle medium-response contributions to jet EECs~\cite{Bossi:2024qho,Yang:2023dwc,Barata:2024ukm,Barata:2025fzd}.

\end{itemize}

\section{Energy Correlators near equilibrium}\label{sec:ETH}
We first explore the collinear angular structure of EECs measured in many-body quantum states. Although the general properties of the celestial OPE for CFTs~\cite{Hofman:2008ar,Kologlu:2019mfz}, and its extension to QCD \cite{Belitsky:2001ij,Chen:2020adz,Chen:2021gdk,Chen:2022jhb}, 
are still valid, their application in this context is far from trivial, and extracting the interesting features of these states using detector operators may require additional tools, as we follow to discuss. 

Let us first briefly review the structure of local operators in many-body quantum systems as postulated by ETH. In its simplest form, ETH states that matrix elements of local operators in the energy--momentum eigenbasis take a universal structure such that, once equilibrium is reached, local observables coincide with their (generalized) canonical values. For a local operator $\mathcal{O}(x)$, which can be expressed through its value at $x=0$ in a translationally invariant system through $\mathcal{O}(x)=e^{ip_\mu x^\mu}\mathcal{O}(0)e^{-ip_\mu x^\mu}$, 
one has
\begin{align}\label{eq:ETH}
\langle a|\mathcal O(0)|b\rangle
= \mathcal O_{\rm th}(0)\,\delta_{ab}
+ e^{-\frac{S(\bar p)}{2}}\,f(\bar p,q)\,R_{ab}\,,
\end{align}
where $|a\rangle$ denotes a high-energy eigenstate that can be labeled by its conserved energy-momentum
$p_a^\mu=(E_a,\vec p_a)$ (and any other conserved charges), and $\bar p^\mu\equiv \tfrac12(p_a^\mu+p_b^\mu)$, $q^\mu = p_b^\mu -p_a^\mu$, and $\mathcal{O}_{th}(0)=\langle \mathcal{O}(0)\rangle_{th}$. 
This equilibrium one-point function $\mathcal{O}_{\rm th}(0)$ is evaluated at the corresponding thermodynamic state fixed by $\bar p^\mu$, while $S(\bar p)$ is the associated entropy. 
The second term describes fluctuations of the system close to equilibrium: $R=R^\dagger$ is a random matrix with statistics,
$\overline{R_{ab}}=0$, $\overline{|R_{ab}|^2}=1$, 
and the envelope function $f(\bar p,q)$ encodes the dependence of off-diagonal matrix elements on the kinematics of the transition, including both the energy difference and momentum transfer. 

Although Eq.~\eqref{eq:ETH} does not by itself fix the structure of generic multi-point correlators, it already constrains lower-point functions; see~\cite{Foini:2018sdb} for generalizations of ETH. In particular, while the long-time value of one-point functions in a thermalizing theory is governed by the diagonal term in Eq.~\eqref{eq:ETH}, the off-diagonal contributions control fluctuations and are directly tied to two-point functions of local operators. Starting from the Lehmann representation of the Wightman function for an eigenstate
\begin{align}
&G^>_{\mathcal{O}\mathcal{O}}(\o,\q)=(2\pi)^4\sum_m |\langle a|\mathcal{O}(0)|m\rangle|^2\delta(\o-(E_m-E_a))\delta^{(3)}(\q-(\p_m-\p_a))\,,
\end{align}
using the ETH postulate and averaging over fluctuations, we find 
\begin{align}
\overline{G}^>_{\mathcal{O}\mathcal{O}}(\o,\q)
\simeq e^{\frac{\beta(p_a)\o}{2}}\left|f\left(p_a+\frac{q}{2},q\right)\right|^2 \, ,
\end{align}
where we have applied that in ETH the number of states in a given energy band is large, such that one can replace the sum by an integral over density of states $\Omega(p_m)\simeq e^{S(E_m)}$, and we have taken $S(E+\o)\simeq S(E)+\beta(E) \o$. Thus, we readily find the ETH realization of the Kubo--Martin--Schwinger (KMS) relation in a translationally invariant system
\begin{align}
&\overline{G}^>_{\mathcal{O}\mathcal{O}}(\o,\q)-\overline{G}^<_{\mathcal{O}\mathcal{O}}(\o,\q)\simeq \left(e^{\frac{\beta(p_a)\o}{2}}-e^{-\frac{\beta(p_a)\o}{2}}\right)\left|f\left(p_a+\frac{q}{2},q\right)\right|^2 \, ,
\end{align}
where we use that $|f\left(\bar{p},q\right)|^2=|f\left(\bar{p},-q\right)|^2$ given $\cO$ is Hermitian and assume that $|f\left(\bar{p}+q/2,q\right)|^2\simeq |f\left(\bar{p}-q/2,q\right)|^2$. Having found the connection between the upper propagator and the form of the fluctuations in $|f|^2$, one can relate the eigenstate Wightman functions with the actual thermal propagators at temperature controlled by $p_a$. To that end, we start with the definition for the thermal Wightman function and apply the ETH prescription:
\begin{align}
G^>_{T}(\o,\q)&=\frac{(2\pi)^4}{Z}\sum_{m,n}e^{-\beta(p_a) E_n}\delta(\o-(E_m-E_n))\delta^{(3)}(\q-(\p_m-\p_n))|\langle n|\mathcal{O}(0)|m\rangle|^2\notag\\
&\simeq e^{\frac{\beta(p_a)\o}{2}} \left|f\left(p_a+\frac{q}{2},q\right)\right|^2\, ,
\end{align}
where we have used a saddle point approximation to integrate over $E_n$ and $\p_n$. One may see that this agrees with $\overline{G}^>_{\mathcal{OO}}$ up to zero frequency modes. It thus follows that the fluctuations can be directly connected with the retarded thermal propagator
\begin{align}\label{eq:f2_GR} 
G_{\cO\cO}^{>}(\omega,\q)
= 
-2(1+n_B(\omega))\mathrm{Im}\,G_T^R(\omega,\q) \, ,
\end{align}
where $q^\mu=p_b^\mu-p_a^\mu$, $\omega=E_b-E_a$, and $n_B(\omega)=\frac{1}{e^{\b\o}-1}$ is the Bose distribution with temperature fixed by $\bar{p}$.

The present discussion, for local operators $\mathcal{O}$, can be extended to light-ray operators~\cite{Delacretaz:2018cfk}. 
To that end let us first decompose the EEC in terms of its disconnected and connected counterparts
\begin{align}
\langle\mathcal{E}(\n_1) \,  \mathcal{E}(\n_2)\rangle = \langle\mathcal{E}(\n_1)\rangle \,  \langle\mathcal{E}(\n_2)\rangle +\langle\delta\mathcal{E}(\n_1) \,  \delta\mathcal{E}(\n_2)\rangle_{\text{conn}} \, ,
\end{align}
where $\delta \mathcal{E}$ denotes the fluctuation of the energy density operator around its expectation value in the given many-body state. If the relative angle is sufficiently large, such that the two points probe two different fluid elements, it is natural to assume that the connected term is suppressed, and the disconnected part dominates the observable. These terms are the ones describing the geometric correlations of the classical (mean) flows, e.g. the contribution of the jet wake~\cite{Barata:2025fzd}. 
Similarly to the case of local operators, the purely connected term carries information about fluctuations around (thermal) equilibrium, which can become more significant as one considers smaller angular separations between detectors, as recently illustrated in the case of a large-charge state in a CFT described by a superfluid EFT~\cite{Cuomo:2025pjp}.

In what follows, we will consider a simple model that allows to generalize the discussion of the connected contribution to the EECs to the case of thermal matter, although relying on multiple additional assumptions and simplifications. Still, the model picture considered here is sufficiently general, indicating that a new angular behavior may emerge in phenomenological considerations. 
We leave the discussion of the uncorrelated terms in analytical hydrodynamic solutions to Section~\ref{sec:Gubser_flow_sec}.

If we assume the medium to be homogeneous and stationary, the purely connected correlator, in analogy with the ETH construction, can be written as
\begin{align}\label{eq:general_connected}
    \langle\mathcal{E}(\n_1) \,  \mathcal{E}(\n_2)\rangle_{\text{conn}}
   & = \lim_{R\to\infty} R^4 \int_0^\infty dt_1 \int_0^\infty dt_2\,\,\n_{1i}\n_{2j}\,\left\langle T_{0i}(t_1,\n_1 R)T_{0j}(t_2,\n_2 R)\right\rangle\notag\\
    &= \lim_{R\to\infty} R^4 \int_0^{\infty} dt_1 \int_0^{\infty} dt_2\,\,\n_{1i}\n_{2j}\int \frac{d^4q}{(2\pi)^4}\,e^{-iq_\m (x_1-x_2)^\mu}G^>_{0i,0j}\left(q\right)\,,
\end{align}
where $G^>_{0i,0j}\left(q\right)$ is the thermal Wightman function of two stress energy tensors. One may notice that if the state is fully time independent, then the resulting correlator is ill-defined, accounting for an infinite-time measurement of the thermal bath fluctuations. However, this issue only appears in a highly idealized formal setting, since
any finite amount of matter
will eventually lose its energy and hadronize, naturally cutting the contributions of the fluctuations at later times. This fact is independent of the specific freeze-out model used here, whose role is only to provide a simple analytic realization of that general point. The contributions of these fluctuations are then carried by nearly free-streaming hadrons to the distant detector. 

\begin{figure}[h]
    \centering
    \includegraphics[width=0.8\linewidth]{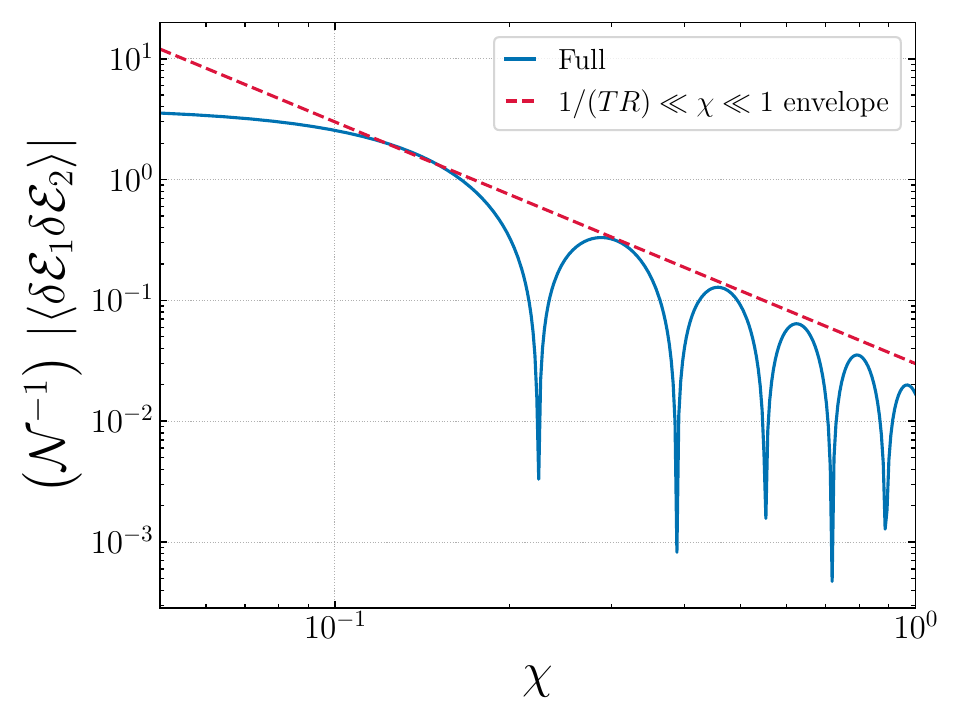}
    \caption{Absolute value of the connected part of $\langle \mathcal{E}(n_1)\mathcal{E}(n_2)\rangle$ divided by its dimensionful pre-factor $\cN = w T^2 R^4/(6\pi^2)$ for the toy model with infinite static thermal matter, hadronizing at a later time. The blue line corresponds to the full result in Eq.~\eqref{eq:inf_full}, while the red dashed line corresponds to the $1/(TR) \ll \chi \ll 1$ envelope, i.e. the $\sim 1/\chi^2$ scaling in Eq.~\eqref{eq:inf_chi2scaling} without the $\cos(\cdot)$. This quantity depends only on two dimensionless parameters: $T R = 20$ and $\tau = T t_{\rm max} = 2$, assuming $c_s=1/\sqrt{3}$.}
    \label{inf_matter_plot}
\end{figure}

It is instructive to illustrate this point by considering a large static volume of the QGP close to the phase transition temperature, immersing the detector of radius $R$ inside it.
In this model, the resulting hadrons are expected to be at rest and will not contribute to the EEC, while the initial static QGP is not flowing through the detector. Thus, the only contribution may come from the fluctuations, and here we focus solely on the hydrodynamic counterpart of these. Substituting the corresponding hydrodynamic Wightman function, see Eq.~\eqref{eq:f2_GR}, we find
\begin{align}\label{eq:inf_starting_point}
\langle\delta\mathcal{E}(\n_1) \,  \delta\mathcal{E}(\n_2)\rangle_{\text{conn}}
&=wR^4\int \frac{d^4q}{(2\pi)^3}\left(\frac{1-\cos \omega t_{\rm max}}{\omega^2}\right)\,e^{-iR\q\cdot(\n_1-\n_2)}\,2(1+n_B(\o))\notag\\
&\hspace{-2cm}\times \left[\frac{\q_i\q_j}{\q^2}\omega^2 {\rm sign}(\omega)\delta(\omega^2 - \q^2 c_s^2)+\left(\delta_{ij}-\frac{\q_i\q_j}{\q^2}\right)\omega\delta(\omega)\right]\n_{1i}\n_{2j} \, ,
\end{align}
where $t_{\rm max}$ is the characteristic time before hadronization, $w$ is the enthalpy of the QGP,  $c_s = 1/\sqrt{3}$ is the speed of sound, and $\omega \equiv q^0$. Since the 
$\omega$ integral is dominated by hydrodynamic poles, we work in the ideal (zero-viscosity) limit to simplify the analysis. Further, we cut the $|\q|$ integrals at the scale of characteristic temperature $T$, since for $|\q|>T$ one expects the hydrodynamic 
discussion to become invalid. Then, performing the $\omega$ and angular integrations, and fixing $\cos\chi = \mathbf n_1 \cdot \mathbf n_2$, one obtains
\begin{align}\label{eq:inf_full}
\notag \langle\delta\mathcal{E}(\n_1)\,&\delta\mathcal{E}(\n_2)\rangle_{\text{conn}} = \frac{w T^2 R^4}{6\pi^2}
\int_0^1 dx \;x^2\Bigg\{\coth\!\left(\frac{c_s x}{2}\right)
\,\frac{1-\cos(c_s \tau x)}{c_s x}\, \bigg[
\cos\chi\, j_0(\lambda x)
\\
&+\frac{3-\cos\chi}{2}\, j_2(\lambda x) \bigg] +\,\tau^2\,\left[ 2\cos\chi\, j_0(\lambda x) -\frac{3-\cos\chi}{2}\, j_2(\lambda x) \right] \Bigg\} \, ,
\end{align}
with $\lambda= 2TR\sin(\chi/2)$, $\tau = Tt_{\rm max}$ and $j_i(\lambda x)$ are spherical Bessel functions. Let us now consider the asymptotic limits of this expression for $\lambda\ll 1$ and $\lambda \gg 1$, assuming $\chi\ll1$. The former regime corresponds to $\chi \ll 1/(TR)$ and yields a $\lambda$-independent result.  
In the latter case, the integrand is bounded and sufficiently regular on the integration domain, uniformly in 
$\lambda$, which allows one to integrate by parts and extract the leading contribution. Hence, for angles $1/(TR) \ll \chi \ll 1$, one finds
\begin{align}\label{eq:inf_chi2scaling}
     \notag \langle\delta\mathcal{E}(\n_1) \,  \delta\mathcal{E}(\n_2)\rangle_{\text{conn}} &\simeq -\frac{T^2R^4w}{2\pi^2\lambda^2}\left[\tau^2\cos^2\frac\chi2-\sin^2\frac\chi2\,\coth\left(\frac{c_s}{2}\right)\frac{1-\cos(c_s\tau)}{c_s}\right]\cos \lambda
     \\
     &\overset{\chi \ll 1}{\to} -\frac{R^2 w \tau^2}{2\pi^2 \chi^2}\cos (TR\chi) \, ,
\end{align}
which scales as $\chi^{-2}$, resembling the behavior of the EEC in the large-charge state~\cite{Cuomo:2025pjp}. The full dependence is shown in Fig.~\ref{inf_matter_plot}. 

While the large-matter, finite-detector-radius setup is illustrative, it corresponds to a state of infinite energy in the limit when $R$ is sent to infinity. To better understand the general features of the EEC in many-body QCD states, it is therefore instructive to consider a model with finite initial energy that evolves into a propagating hadronic state. Such state can then reach an arbitrarily distant detector, carrying imprints of the fluctuating matter. Let us then start by considering an equilibrated spherical (toy) QGP shell with radius $r_{\rm in}$, thickness $d$, and at a temperature close to the hadronization phase transition, maintained in a stationary state by external sources that balance its expansion. At $t=0$ the shell starts evaporating from the outer surface. To proceed, we assume that hadronization is sufficiently local and rapid, neglecting both hadronic interactions and hydrodynamic expansion. 
Under these assumptions, we express the resulting stress-energy tensor in terms of the corresponding initial conditions just prior to hadronization, namely the stress-energy tensor in the QGP phase.
Consequently, correlators of the stress-energy tensor at a distant detector can be related to the corresponding correlators near the hadronization surface. Assuming the outgoing hadronic shell propagates at the speed of light, this leads to
\begin{align}
T^{\mu\nu}(t, R\n)=\frac{1}{R^2}l^\mu(\n)l^\nu(\n)\frac{r_s^2\left(t_e\right) P(t_e,r_s(t_e))}{1+v_s}\theta(t_e)\theta\left(d-v_s t_e\right)\Bigg|_{t_e=\frac{t-R+r_{\rm in}}{1+v_s}}\, ,
\end{align}
where $l^{\mu}(\mathbf n) = (1,\mathbf n)$, and we assume that the hadronization surface propagates at constant speed from the outer to the inner boundary of the shell, $r_s(t) = r_{\rm in} - v_s t$, with $v_s = P/\epsilon$. Focusing on the limit $R\gg r_{\rm in}$ and $r_{\rm in}\gg d$, we readily find that
\begin{align}
\mathcal{E}(\n) = \frac{1}{3}\int^{t_{\rm max}}_0 dt_h r_s^2(t_h)\epsilon(t_h, r_s(t_h)\n)\, ,
\end{align}
%
where $t_{\rm max} = d/v_s$, and we use the conformal equation of state. The resulting connected term in the EEC can be written as 
\begin{align}
\langle\delta\mathcal{E}(\n_1) \,  \delta\mathcal{E}(\n_2)\rangle_{\text{conn}} = \frac{1}{9}\int^{t_{\rm max}}_0dt_1\int^{t_{\rm max}}_0dt_2\,r_s^2(t_1)r_s^2(t_2)\,\langle\delta\epsilon(t_1,r_s(t_1)\n_1)\delta\epsilon(t_2,r_s(t_2)\n_2)\rangle \,.
\end{align}

\begin{figure}[b]
    \centering
    \includegraphics[width=0.8\linewidth]{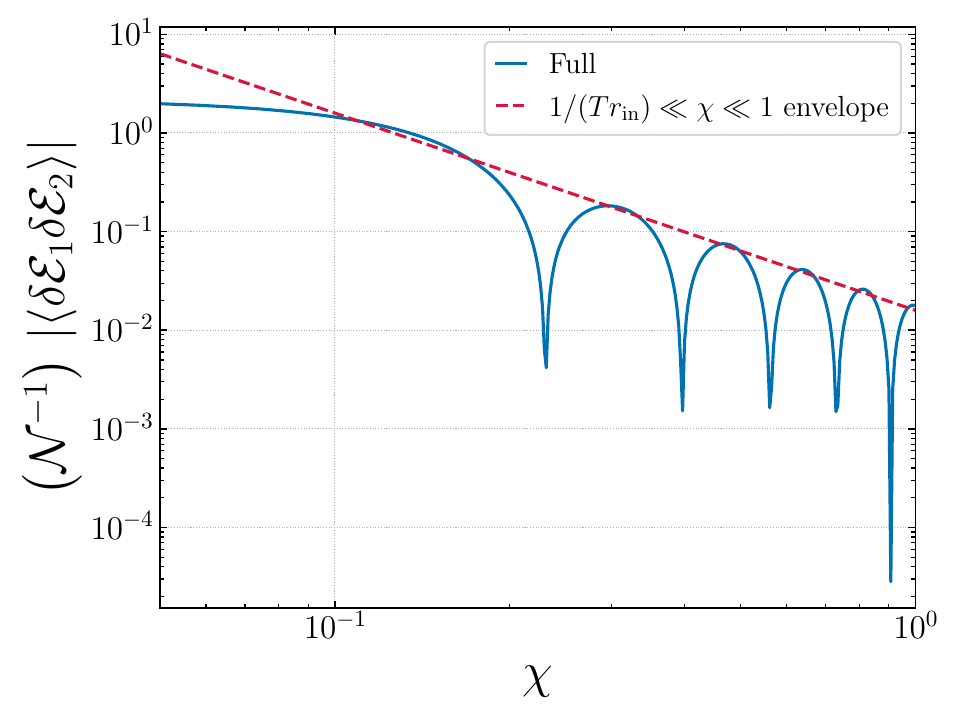}
    \caption{Absolute value of the connected part of $\langle \mathcal{E}(n_1)\mathcal{E}(n_2)\rangle$ divided by its dimensionful pre-factor $\cN = w \tau T^2 r^4_{\rm in}/(36\pi^2)$ for the toy model with a finite hadronizing shell as a function of $\chi$. The blue line corresponds to the full result in Eq.~\eqref{eq:shell_full}, while the red dashed line corresponds to the $1/(Tr_{\rm in}) \ll \chi \ll 1$ envelope, i.e. the $\sim 1/\chi^2$ scaling in Eq.~\eqref{eq:shell_chi2scaling} without the $\cos(\cdot)$. This quantity depends only on two dimensionless parameters which we set as $T r_{\rm in} = 20$ and $\tau = T t_{\rm max} = 2$, assuming $c_s=1/\sqrt{3}$ and $v_s=1/3$. This implies that $T d = v_s \tau = 2/3$, so for $\chi \ll 1$ we are effectively in the regime $\chi \ll 1/(T d)$.}
    \label{shell_plot}
\end{figure}

While ETH motivates the expectation that local fluctuations in a near-equilibrium many-body state are governed by thermal or hydrodynamic two-point functions, for a finite expanding shell the relevant description is given by the Wightman stress-tensor correlator in a locally equilibrated background. We therefore take this as our starting point. Using the Wightman function for the hydrodynamic fluctuations in the ideal limit and approximating $r_s(t)\simeq r_{\rm in} $ everywhere except in the phases we find
\begin{align}
&\langle\delta\mathcal{E}(\n_1) \,  \delta\mathcal{E}(\n_2)\rangle_{\text{conn}} \simeq \frac{1}{9}wr_{\rm in}^4\int dt_1dt_2\,\int\frac{d^4q}{(2\pi)^3}e^{-i\q\cdot\left[r_s(t_1)\n_1-r_s(t_2)\n_2\right]}e^{i\o(t_1-t_2)}\notag\\
&\hspace{5cm}\times \q^2(1+n_B(\omega))\text{sign}(\o)\delta\left(\omega^2-\q^2c_s^2\right) \, .
\end{align}
Once again applying a cutoff $|\q| <T$ 
we arrive at
\begin{align}\label{eq:shell_full}
    & \langle\delta\mathcal{E}(\n_1) \,  \delta\mathcal{E}(\n_2)\rangle_{\text{conn}} \simeq \frac{T^2\,\tau w r_{\rm in}^4}{36\pi^2 c_s}\int_0^{1} dy\int_0^1 dx\,x^2\int_{-1}^{1} du\;\,J_0\left(\zeta \left(1 - \frac{v_s t_{\rm max}}{r_{\rm in}}y\right)\,x\sqrt{1-u^2}\right)\nn
    &\hspace{3cm}\times\coth\left(\frac{c_s x}{2}\right)\frac{\sin\left((c_s-v_s u\cos\frac{\chi}{2})(x\,\tau)\,f(y)\right)}{c_s-v_s u\cos\frac{\chi}{2}}\,.
\end{align}
where
$\tau = Tt_{\rm max}$, $f(y) = 2\,{\rm min}(y, 1-y)$ and $\zeta = 2Tr_{\rm in}\sin\chi/2$ with rescaled integration variables $x=q/T$ and $y = (t_1+t_2)/2 t_{\rm max}$.
Similarly to the case of static matter, 
the connected part of the correlator is $\zeta$-independent for $\chi \ll 1/(T r_{\rm in})\ll1$, or equivalently for $\zeta \ll 1$. For $1/(T r_{\rm in})\ll \chi\ll1$, the oscillatory behavior of the Bessel function implies that the dominant contribution to the $u$ integral comes from $u \sim 0$. Since the $\zeta$-independent factors are regular in $u$ and $x$, one can integrate by parts in $x$ to extract the leading contribution. Provided the finite thickness of the emitting shell is negligible at the scale set by the angular phase, namely $\zeta v_st_{\rm max}/r_{\rm in} \sim T d\,\chi\ll 1$, one may neglect the $y$-dependence of the Bessel phase at leading order. Thus, for angles $1/(T r_{\rm in}) \ll \chi \ll 1/(Td)$, we arrive at\footnote{
The $\chi^{-2}$ scaling holds as long as the angular resolution is insufficient to resolve the thickness of the emitting shell. For larger angles, $T d\,\chi \gtrsim 1$, there is additional suppression, and the asymptotic behavior crosses over to $\chi^{-4}$.}
\begin{align}\label{eq:shell_chi2scaling}
    \langle\delta\mathcal{E}(\n_1) \,  \delta\mathcal{E}(\n_2)\rangle_{\text{conn}} =- \frac{w r_{\rm in}^2}{9 \pi^2 c_s^3}\frac{\cos\left(r_{\rm in} T \chi\right)}{\chi^2}\coth\frac{c_s}{2}\sin^2\left(\frac{c_s \tau}{2}\right) \, ,
\end{align}
which, as in the previous case, scales as $\chi^{-2}$. The full behavior of the EEC in this state is shown in Fig.~\ref{shell_plot}, exhibiting a structure similar to that found in the previous model.

The results of the previous two toy models involve a hard UV-dependent cutoff at the temperature scale, which leads to the overall cosine scaling. It is natural to wonder wether this UV-dependent behavior modifies in the full UV-complete theory. 
It is thus instructive to consider the opposite regime of large $\omega$ and $|\q|$. 
We start by taking the kinetic theory two-point function, applicable well beyond the hydrodynamic regime, see e.g. \cite{Romatschke:2015gic}, in the large relaxation time limit
\begin{align}
&G_{0i,0j}^>(\omega,\q)=3\pi w (1+n_B(\omega))\notag\\
&\hspace{4cm}\times\left[\frac{\q_i\q_j}{\q^2}\frac{\omega^3}{|\q|^3}+\left(\delta_{ij}-\frac{\q_i\q_j}{\q^2}
\right)\frac{\omega(\q^2-\omega^2)}{2|\q|^3}\right]\Theta(|\q|-|\omega|) \, ,
\end{align}
and consider the EEC in the case of a large static volume of matter, similarly to the first toy model above.
Focusing on the asymptotic scaling for $|\q|\gg T$ and working in the small angle limit $\chi\ll1$, after subtracting the zero temperature 
contribution, we find that the connected part of the EEC  
reads
\begin{align}
\langle\delta\mathcal{E}(\n_1) \,  \delta\mathcal{E}(\n_2)\rangle_{\text{conn}} &\propto wR^4\int\frac{d^3\q}{(2\pi)^3}\int_{0}^{|\q|} d\omega\, \left(\frac{1-\cos \omega t_{\rm max}}{\omega^2}\right)e^{-i |\q| R \cos\theta \chi}\nn
&\hspace{0cm}\times \left(\coth\frac{\omega}{2T}-1\right)\left(\frac{\q_i\q_j}{\q^2}\frac{\omega^3}{|\q|^3}+\left(\delta_{ij}-\frac{\q_i\q_j}{\q^2}\right)\frac{\omega(\q^2-\omega^2)}{2|\q|^3}\right)\n_{1i} \n_{2j} \,.
\end{align}
Taking the frequency integral, we notice that it saturates at $\omega\sim T$, resulting in
\begin{align}\label{eq:KT_scaling}
&\langle\delta\mathcal{E}(\n_1) \,  \delta\mathcal{E}(\n_2)\rangle_{\text{conn}} \propto \frac{wR^2}{\chi^2} \, \int\limits^{\chi \Lambda_{\rm UV} R} dx \sin x  \, , 
\end{align}
for $\chi \gg 1/(R\, \Lambda_{\rm UV})$, where the momentum integral is now regulated at some UV scale $\Lambda_{\rm UV}$ and not at the temperature. 
The oscillating behavior in $\chi$ is still present, 
as a consequence of 
the result's UV dependence. 
In the complete theory, one expects this feature to be regulated by a smooth transition to the UV regime. It should also be mentioned here that 
the contributions at smaller momenta and frequencies will result in the associated hydrodynamic scaling omitted here.

Combining these considerations with the usual small angle structure, we are lead to the generic angular scaling already illustrated in Fig.~\ref{fig:cartoon}. There, we only indicate the leading scaling behavior, while implicit subleading terms can become more relevant in the corresponding limits, e.g. the OPE at larger separations. At large angles, there is a linear rise coming from the uncorrelated contributions, which is universal. As the angle decreases the connected terms become prevalent, indicating a departure from an (ideal) gas behavior. This evolution is determined by the hydrodynamic modes. 
However, the full structure, as illustrated in Figs.~\ref{inf_matter_plot},~\ref{shell_plot} ,
is harder to constrain in general, since it depends on the details of the full microscopic theory of the matter under consideration. Nonetheless, we argue that this behavior should be observed even in the case of, e.g., finite coupling $N=4$ SYM, where in principle this transition should appear. As the angle decreases, the hydrodynamic description becomes inapplicable, at which point the EEC must be matched onto the OPE beyond leading twist. 
We indicate the transition by a gray band, as we do not describe it. The characteristic angle $\mu^{-1}\sim (Tr_{\rm ch})^{-1}$, is tied to the global properties of the matter state.

\section{Geometric correlations in EECs in boost invariant flowing matter}\label{sec:Gubser_flow_sec}

In this section, we turn to the disconnected part of the EECs governed by the flow of the QCD matter produced in HICs during its hydrodynamic QGP phase.
As has been mentioned above, the many-body phase of the QCD matter produced in HIC cannot be probed directly. Any phenomenologically relevant consideration of the EECs in HICs has to focus on the final state particles, reaching a distant detector.
Crucially, these hadrons carry the imprint of the many-body evolution of the system. Here, we consider how the details of the mean hydrodynamic evolution during the QGP phase translate into the final state EECs.

\subsection{EECs of a boost-invariant flow}

We start by focusing on the hadronization process and the corresponding phase transition hypersurface, separating the QGP phase from the vacuum. Following the logic of the Cooper-Frye prescription for the hadronization transition~\cite{PhysRevD.10.186,Cooper:1974mv}, see e.g.~\cite{Ivanov:2008zi} for review of alternative hadronization strategies, we notice that at this hypersurface the stress energy tensor has to stay continuous along the normal
\begin{align}\label{eq:matching_fluid_hadron}
	 \xi_\mu(x_h)  T^{\mu \nu}_{\rm fluid}(x_h)  = \xi_\mu(x_h) T^{\mu \nu}_{\rm had.}(x_h)\,,
\end{align}
where $\xi_\mu \equiv -\partial_\mu \epsilon$ is the hypersurface outward-pointing normal vector for $x_h$ satisfying $\epsilon(x_h)=\epsilon_h$, and we assume that there is a single flavor of hadrons of mass $m_h$ for simplicity. Keeping the hydrodynamic side unspecified for the moment, we notice that for a gas of hadrons the stress energy tensor is fully fixed by the kinetic theory prescription
\begin{align}\label{eq:EMT_kinetic_theory}
	T_{\rm had.}^{\mu\nu}(x) = \int\frac{d^3 p}{p^0}p^\mu p^\nu f(x,p)\,,
\end{align}
where $f(x,p)$ is the distribution of the hadrons in the final state and $p^2=-m_h^2$. Ignoring rescatterings, one may further assume that the hadrons are free-streaming and the distribution at large distance/late time is fully fixed by the initial distribution at the hadronizaton hypersurface. Thus, Eq.~\eqref{eq:matching_fluid_hadron} may allow us to relate the information on the hydrodynamic evolution with the EECs built on the final state particles.

For a sufficiently local hadronization process, we assume that a hydrodynamic cell on the freeze-out surface at point $x_h$ gives rise to a density of hadrons $n(x_h)$ with a fixed momentum $p_h(x_h)$, defined by the continuity condition in Eq.~\eqref{eq:matching_fluid_hadron}. Then, the initial distribution at $x_h$ can be written as $f(x_h,p) = n(x_h) \delta^{(3)}(\vec p - \vec p_h(x_h))\Theta(p_h\cdot \xi)$, where we only consider the outgoing part of the distribution ($p_h\cdot \xi>0$).\footnote{In choosing $p_h\cdot \xi>0$ contributions only, we are interpreting the freeze-out hypersurface as converting fluid degrees of freedom into outgoing, free-streaming hadrons, i.e., we do not model additional physical mechanisms accounting for reabsorption, back-reaction, \textit{etc}.} Following the hadrons along their straight trajectories, one may readily express the distribution through its values on the hadronization hypersurface
\begin{align}
	f(x,p) = \int d^4 x_h \delta(\epsilon(x_h) - \epsilon_h)\frac{|p\cdot \xi(x_h)|}{p^0}\Theta(t-t_h)\delta^{(3)}\left(\vec x - \vec{x}_h  - \frac{\vec p}{p^0}(t-t_h)\right)f(x_h,p)\,,
\end{align}
where one should pay particular attention to the Jacobian, which is required for consistency at the phase surface. Naturally, this distribution solves the collisionless Boltzmann equation $p^\mu \partial_\mu f(x,p) = 0$, being a function of $\vec{x}-(\vec{p}/p^0)t$ only, and closely follows the commonly used Cooper-Frye prescription.

At the freeze-out surface the hadronic stress energy tensor takes a particularly simple form and the continuity relation in Eq.~\eqref{eq:matching_fluid_hadron} reads
\begin{align}\label{eq:matching_explicit}
    &A^\nu(x_h) \equiv \xi_\mu(x_h)T^{\mu \nu}_{\rm fluid}(x_h)= n(x_h)\Theta(p_h\cdot \xi)\frac{(p_h(x_h)\cdot \xi(x_h))p_h^\nu(x_h)}{p_h^0(x_h)}\,.
\end{align}
Thus, the hadron momentum at the hadronization surface is fixed by $p^\mu(x_h) / p_h^0 = A^\mu(x_h) / A^0$, noticing that $A^0>0$. The resulting hadronic stress energy tensor is independent of $m_h$ and $n(x_h)$, and can be written as
\begin{align}
\label{eq:final_hadronic_EMT}
	\notag T_{\rm had.}^{\mu\nu}(x) &= \int d^4 x_h\, \delta(\epsilon(x_h)-\epsilon_h)\Theta(A^0(x_h)) \frac{A^\mu(x_h) A^\nu(x_h)}{A^0(x_h)}
    \nn
    &\times
    \Theta(t-t_h)\delta^{(3)}\Big(\vec x -\vec{x}_h - \frac{\vec A(x_h) }{A^0(x_h)}(t-t_h)\Big)\,.
\end{align}

Before turning to the specific HIC motivated hydrodynamic profile, it is also instructive to consider the generic form of the energy flow operator of the final state hadrons. Assuming that the initial QGP droplet is localized and fully hadronized over a short time\footnote{One should notice that when we later consider the Gubser flow in Section~\ref{sec:Gubser_flow_sec}, this assumption breaks. However, by imposing additional cuts in the EEC, it can be fully recovered.} (its spacetime dimensions are much smaller than the characteristic size of the detector), we can write 
\begin{align}
\label{eq:EPS_before_fluid}
	\mathcal{E}(\vec n)  = \int d^4 x_h\, \delta(\epsilon(x_h)-\epsilon_h)\Theta(A^0(x_h))A^0(x_h)\delta^{(2)}(\Omega_{\vec v}-\Omega_{\vec n}) \, ,
\end{align}
where $\vec v(x_h) \equiv \vec A(x_h) / A^0(x_h)$. We will focus on a simple analytic boost-invariant flow of a viscous conformal fluid, having in mind the analytic case of Gubser flow~\cite{Gubser:2010ui}.  
We further assume that azimuthal perturbations to the flow are small, effectively expanding around a rotationally symmetric configuration. 
Parameterizing $A^{\mu}$ in Milne coordinates, where $ds^2 = -d\tau^2 + dx_\perp^2 + x_\perp^2 d\phi^2 + \tau^2 d\eta^2$, one may further
write the energy flux as
\begin{align}
	\mathcal{E}(\vec n)  = \int \tau d\tau x_\perp d x_\perp d\phi\, \delta(\epsilon(\tau, x_\perp,\phi)-\epsilon_h)\Theta(A^\tau)A^\tau(\tau, x_\perp, \phi)\int d\eta \cosh\eta\,\delta^{(2)}(\Omega_{\vec v}-\Omega_{\vec n}) \,,
\end{align}
where we have omitted subscript $h$ on the Milne coordinates for compactness. At leading order in the azimuthal perturbations and in the corresponding domain, we can further write
\begin{align}
&\int d\eta\cosh\eta\,\delta^{(2)}(\Omega_{\vec v}-\Omega_{\vec n}) 
\simeq\frac{1}{\sin^3\theta_\bn}\delta(\phi-\phi_\bn)\left\{1-\frac{\pa_\phi A^{\phi}}{A_{\perp}}x_{\perp}\right\}|A_\perp/A^\tau|\,,
\end{align}
arriving at a particularly compact form for the energy flux operator of boost invariant flowing matter
\begin{align}\label{eq:EFO_general}
	\mathcal{E}(\theta_\bn,\phi_\bn)  = \frac{\cF(\phi_\bn)}{\sin^3\theta_\bn}\,,
\end{align}
where 
\begin{align}\label{eq:EFO_coefficient}
	\cF = \int_0^\infty \tau d\tau \int_0^\infty x_\perp d x_\perp\, \delta(\epsilon(\tau, x_\perp,\phi_\bn)-\epsilon_h)\Theta(A^\tau) 
     |A_\perp(\tau, x_\perp,\phi_\bn)|\left\{1-\frac{\pa_\phi A^{\phi}}{A_{\perp}}x_{\perp}\right\}\Bigg|_{\phi=\phi_\bn}\,.
\end{align}

\subsection{Probing Azimuthal Perturbations}\label{subsec:azimuthal_EEC}

With this general form of the energy flow operator for hadrons emitted off a boost-invariant hydrodynamic flow in hand, we can now consider how its details are imprinted in projected EECs defined as
\begin{align}\label{eq:azimuthal_EEC_def}
    &\frac{d\Sigma^\perp}{d\cos\chi\, d\cos\psi} =
    {\rm sgn}(\sin\psi)\int_{\n_1,\n_2}
    \frac{\mathcal E(\n_1)\mathcal E(\n_2)}{Q^2}\,
    \delta(\n_1\!\cdot\!\n_2-\cos\chi)\notag\\
    &\hspace{5cm}\times \delta\!\left(\n_{s,\perp}\cdot \hat{\boldsymbol{x}}
    -\cos\psi\right)\Theta\left(\n_{s,\perp}\cdot \hat{\boldsymbol{y}}\sin\psi\right)C_\Delta(\theta_1,\theta_2)\,,
\end{align}
where the sgn$(\sin\psi)$ and $\Theta\left(\n_{s,\perp}\cdot \hat{\boldsymbol{y}}\sin\psi\right)$ allow one to extend the range to $\psi\in[0,2\pi]$, $\hat{\boldsymbol{x}}$ and $\hat{\boldsymbol{y}}$ are fixed perpendicular directions in the plane transverse to the beam, $\n_{s,\perp} = \frac{\n_s - (\n_s\cdot \boldsymbol{\hat z})\boldsymbol{\hat z}}{\sqrt{1- (\n_s\cdot \boldsymbol{\hat z})^2}}$ is the normalized projection of $\n_s = (\n_1+\n_2)/|\n_1+\n_2|$, and $Q$ is some reference energy scale. The function $C_\Delta(\theta_1, \theta_2)=\Theta(\cos\Delta-|\cos\theta_1|)\Theta(\cos\Delta-|\cos\theta_2|)$ applies a cut in $\theta_1$ and $\theta_2$, focusing the EEC on the region near mid-rapidity and removing the divergent forward and backward flows of the boost-invariant solution. 

Let us now write $\n_1$ and $\n_2$ in terms of $\n_s$, the angle $\gamma$ between $\n_1$ and $\n_2$ and $\phi_d$, the azimuthal angle of their difference $\n_d = (\n_1-\n_2)/|\n_1-\n_2|$ around $\n_s$. To do so, we first recall that $\n_s\cdot\n_d = 0$, which constrains these unit vectors and allows to define $\hat{\n}_d = (\cos\phi_d,\sin\phi_d,0)$ in the frame where $\n_s ||\, \hat{e}_z$. This set of three independent angles should be complemented with $\cos\gamma \equiv \n_1\cdot\n_2$, yielding the following expression for the EEC integrals 
\begin{align}
   \notag \frac{d\Sigma^\perp}{d\chi d\psi}&=\sin\chi\int\, d^2\Omega_s\,d\phi_d\, d\cos\gamma\,\frac{\mathcal{E}(\theta_1,\phi_1)\mathcal{E}(\theta_2,\phi_2)}{Q^2}
    \\
    &\hspace{1cm}\times \delta(\cos\gamma-\cos\chi)\delta(\phi_s-\psi)C_\Delta(\theta_1,\theta_2)\,. 
\end{align}
Expanding the energy flux expression in Eq.~\eqref{eq:EFO_general} in azimuthal perturbations as $\cF = \sum_n\left[\varepsilon_n \cos n\phi \,\cF^c_n+\delta_n \sin n\phi \,\cF^s_n\right]$, with $\varepsilon_0=1$, we find the following expression for the projected EEC
\begin{align}
    \frac{d\Sigma^\perp}{d\chi d\psi} = \frac{\cF_0}{Q^2}\left(\frac{\cF_0f_0(\chi)}{2} + \sum_{n\geq 1}
     f_n(\chi)\big[\varepsilon_n \cF^c_n\cos n\psi +\delta_n \cF^s_n\sin n\psi \big]\right)\,,
\end{align}
where
\begin{align}
    &f_n(\chi) = 2\sin\chi \int_{-1}^1 d\cos\theta_s \,\int_0^{2\pi} d\phi_d \frac{\cos n \delta}{\sin^3\theta_1\sin^3\theta_2} C_\Delta(\theta_1,\theta_2)\, ,
\end{align}
with
\begin{align}
    & \cos\theta_{1,2} = \cos(\chi/2)\cos\theta_s \mp \sin(\chi/2)\sin\theta_s\cos\phi_d\,,\nn
    & \tan\delta = \frac{\sin(\chi/2)\sin\phi_d}{\cos(\chi/2)\sin\theta_s+\sin(\chi/2)\cos\phi_d \cos\theta_s}\,.
\end{align}
We find that a small $\chi$ expansion of this function reads $f_n(\chi) = \alpha_0\chi + \alpha_1 |\chi|\chi + \alpha_{2}\chi^3 + \mathcal{O}(\chi^4)$, where the leading contribution is linear in $\chi$, in agreement with~\cite{Barata:2024ukm,Barata:2025fzd}, while the first $n$-dependent term enters through $\alpha_2$.

To further explore the azimuthal asymmetry of this projected EEC, we adapt an approach commonly used in studies of the azimuthal structure of the flowing QGP medium \cite{Voloshin:1994mz,Voloshin:2008dg,Heinz:2013th} and define the EEC harmonics\footnote{It is worth noting in passing that, at linear order in perturbations, the EEC harmonics are in some respects similar to the energy-weighted harmonics, see e.g. \cite{Liu:2015nwa,Kurkela:2018ygx,Kurkela:2019kip}.} as (see e.g.~\cite{Barata:2023zqg,Barata:2024bqp,Barata:2025uxp}) 
\begin{align}\label{eq:EEC_harmonics}
      & v^{\rm EEC}_n = \frac{1}{\pi}\int_0^{2\pi} d\psi \cos n\psi \frac{d\Sigma^\perp}{d\chi d\psi} =  \frac{\cF_0^2}{Q^2}\,\frac{\varepsilon_n \cF^c_n}{\cF_0}f_n(\chi) \, ,
      \notag\\
      & w^{\rm EEC}_n = \frac{1}{\pi}\int_0^{2\pi} d\psi \sin n\psi \frac{d\Sigma^\perp}{d\chi d\psi} =  \frac{\cF_0^2}{Q^2}\,\frac{\delta_n \cF^s_n}{\cF_0}f_n(\chi)\,.
\end{align}
This allows us to
quantify the azimuthal asymmetry
directly from the perturbations to the hydrodynamic solution, with the magnitude of the initial perturbations factored out. For instance, in the case of cosine terms, we find
\begin{align}\label{eq:EEC_harmonics_ratio}
    \frac{1}{\varepsilon_n}\left(\frac{v_n^{\rm EEC}}{v_0^{\rm EEC}}\right) = \frac{\cF^c_n}{\cF_0}\frac{f_n(\chi)}{f_0(\chi)} = \frac{\cF^c_n}{\cF_0}\left(1 - \frac{n^2}{16}\,\left(I_4/I_3\right)\chi^2+\mathcal{O}(\chi^3)\right)\,,
\end{align}
where we have defined $I_m=\int^{\cos\Delta}_{-\cos\Delta}\frac{dx}{(1-x^2)^m}$. This relation quantifies the azimuthal structure of energy correlations in terms of geometric coefficients $f_n(\chi)$ and $\cF_n^c$. Near mid-rapidity $\Delta \sim \pi/2$ this result has only a mild dependence on the cut, i.e., $I_4/I_3 = 1 + \cO((\Delta-\pi/2)^2)$. Thus, one may see that the azimuthal perturbations of a specific flow, encoded in $\cF_n^{c/s}$, can be accessed through the azimuthally differential EECs.

\subsection{Gubser Flow}
To provide a quantitative picture of obtained results, we further consider a specific boost-invariant hydrodynamic solution, the Gubser flow~\cite{Gubser:2010ui,Gubser:2010ze} for a conformal fluid, $\epsilon=3P$. Its thermodynamic properties are fixed by
\begin{align}\label{eq:Gubser_energy_density}
\epsilon=\frac{\hat{\epsilon}(g)}{\tau^4}\equiv\frac{1}{\tau^4}\left[\frac{\hat{T}_0}{(1+g^2)^{1/3}}+\frac{H_0g}{(1+g^2)^{1/2}}\left\{1-(1+g^2)^{1/6}{}_2F_1\left(\frac{1}{2},\frac{1}{6};\frac{3}{2};-g^2\right)\right\}\right]^4 \, ,
\end{align}
where $\tau=\sqrt{t^2-z^2}$ is the proper time, $g=\frac{1-q^2(\tau^2-x_\perp^2)}{2q\tau}$, $\hat{T}_0$ is a free parameter, and $H_0$ controls viscosity via $\eta=H_0\epsilon^{3/4}$. In turn, the corresponding velocity field is given by
\begin{align}
u^\mu=\left\{u^\tau,u^\perp,u^\phi,u^\eta\right\}=\left\{\cosh\kappa,\sinh\kappa,0,0\right\} \, ,
\end{align}
where $\tanh\kappa=\frac{2q^2\tau x_\perp}{1+q^2\tau^2+q^2x_\perp^2}$. 

The expectation of the energy flux operator on this state can be now expressed through the hydrodynamic stress energy tensor, following Eqs. \eqref{eq:matching_explicit}, \eqref{eq:EFO_general}, and \eqref{eq:EFO_coefficient}. At the first order in hydrodynamic expansion and in the Landau frame, where the Gubser flow is commonly defined, it is given by
\begin{align}\label{eq:FluidTensor}
T^{\mu\nu}&=wu^\mu u^\nu + p g^{\mu\nu} - \eta \sigma^{\mu\nu}\,,
\end{align}
where $\sigma^{\mu\nu}=P^{\mu\alpha}P^{\nu\beta}\left(\nabla_{\alpha}u_{\beta}+\nabla_{\beta}u_{\alpha}-\frac{2}{3}g_{\alpha \beta}\nabla\cdot u\right)$ and $P^{\mu\nu}=g^{\mu\nu}+u^\mu u^\nu$, and $u^2=-1$ in our notation. The hadronization surface is defined by $\hat{\epsilon}(g)/\tau^4=\epsilon_h$, which determines the proper time $\tau=\tau_h(g)$ as a function of $g$. Its normal reads
\begin{align}
\xi_\mu=\frac{\epsilon_h}{\tau}\left(4\d_\mu ^\tau-\frac{\hat{\epsilon}'(g)}{\epsilon_h\tau^4}u_\mu(\tau,g)\sqrt{1+g^2}\right)\,.
\end{align}
Thus, the relevant components of the projection of the stress energy tensor at the hadronization surface are given by
\begin{align}\label{eq:AGeneral_unperturbed}
&A^\tau  = 
-\frac{4}{3}\frac{q\epsilon_h^2}{(q\tau)^2(1+g^2)^{3/2}}\Bigg[\frac{q H_0}{\epsilon_h^{1/4}}\,g\left(2+g^2-3g(q\tau)-2(q\tau)^2\right)\nn
&\hspace{3cm}+\sqrt{1+g^2}\,(q\tau)\left(1-g^2-6g(q\tau)-4(q\tau)^2\right)\Bigg]\,,\nn
& A_\perp = 
\frac{4}{3}\frac{q\epsilon_h^2}{(q\tau)^2(1+g^2)^{3/2}}(g+2(q\tau))\left(\frac{qH_0}{\epsilon_h^{1/4}}g+2(q\tau)\sqrt{1+g^2}\right)(qx_{\perp})\,,
\end{align}
Changing integration variables from $(\tau,x_\perp)$ to $(\tau, g)$ through $(q x_\perp)^2=2(q\tau)g+(q\tau)^2-1$ in Eq.~\eqref{eq:EFO_coefficient} and integrating over $\tau$ to set $\tau = \tau_h(g) = (\hat \epsilon(g)/\epsilon_h)^{1/4}$, we arrive at
\begin{align}\label{eq:F_simp}
    \cF = \frac{\epsilon_h}{q^3}\int_{-\infty}^{+\infty} dg\,\Bigg[\frac{(q\tau)^3}{4}\,\Theta(x_\perp^2(\tau,g))\Theta(A^\tau(\tau,g))\,\frac{| A_\perp(\tau,g)|}{q\epsilon_h^2}\Bigg]_{\tau = \tau_h(g)}\,,
\end{align}
where we have factored out the dimensionful pre-factor $\epsilon_h/q^3$. One can show that the integrand is then a function of $R_\epsilon^{1/4} = \epsilon_h^{1/4}/\left(q\hat T_0\right)$, with $\hat T_0$ the free parameter entering Eq.~\eqref{eq:Gubser_energy_density}, and of $\bar H_0 = H_0/\hat T_0$. To settle on physically meaningful values for these parameters, we follow~\cite{Gubser:2010ui,Gubser:2010ze}, such that we set energy density at freeze-out as $\epsilon_h \simeq 11 T_c^4 \sim 0.83\, {\rm GeV}/{\rm fm}^3$ for $T_c \sim 0.155$ GeV, as commonly used in hydrodynamic simulations of HICs~\cite{Moreland:2014oya,HotQCD:2014kol}, the free parameter to $\hat T_0 = 5.55$, matching to the final entropy per unit rapidity,\footnote{Note that for phenomenological applications one should choose a different value for $\hat T_0$ depending on viscosity $H_0$, since entropy is produced during the hydrodynamic evolution for non-zero $H_0$, see~\cite{Gubser:2010ui,Gubser:2010ze}.} and $q^{-1} \in (4,10)$ fm for typical values of transverse size of the system. We further consider two setups with $\eta=0$ (inviscid) and varying $\eta \neq 0$ (viscous). Given the characteristic value of $\eta/s=1/(4\pi)$ corresponding to $H_0 = 0.33$, we shall vary $H_0 \in (0.05,1)$ for the viscous analysis.

\begin{figure}[h!]
    \centering
    \includegraphics[width=0.75\linewidth]{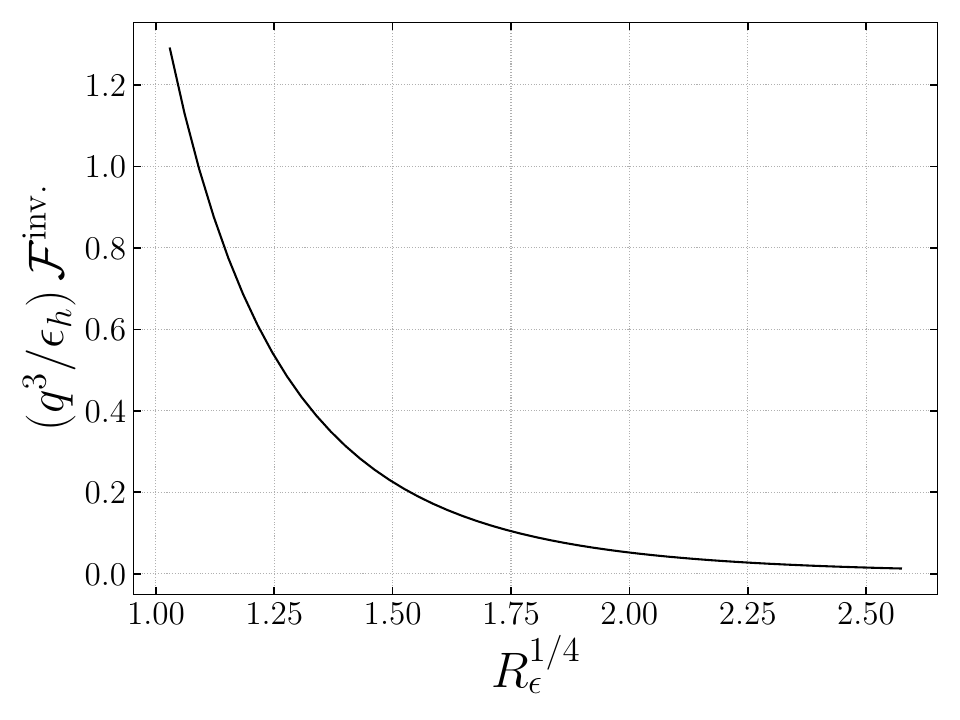}
    \caption{$\cF$ defined in Eq.~\eqref{eq:F_simp}, for zero viscosity flow, divided by its dimensionful pre-factor and as a function of the dimensionless parameter $R_\epsilon^{1/4} = \epsilon_h/\left(q\hat T_0\right)$. The range corresponds to $q^{-1} \in (4,10)$ fm, $\hat T_0 \simeq 5.5$ and $\epsilon_h \simeq 0.83$ GeV/fm$^3$.}
    \label{fig:Fbar_vs_invq_inviscid}
\end{figure}

\begin{figure}[h!]
    \centering
    \includegraphics[width=0.45\linewidth]{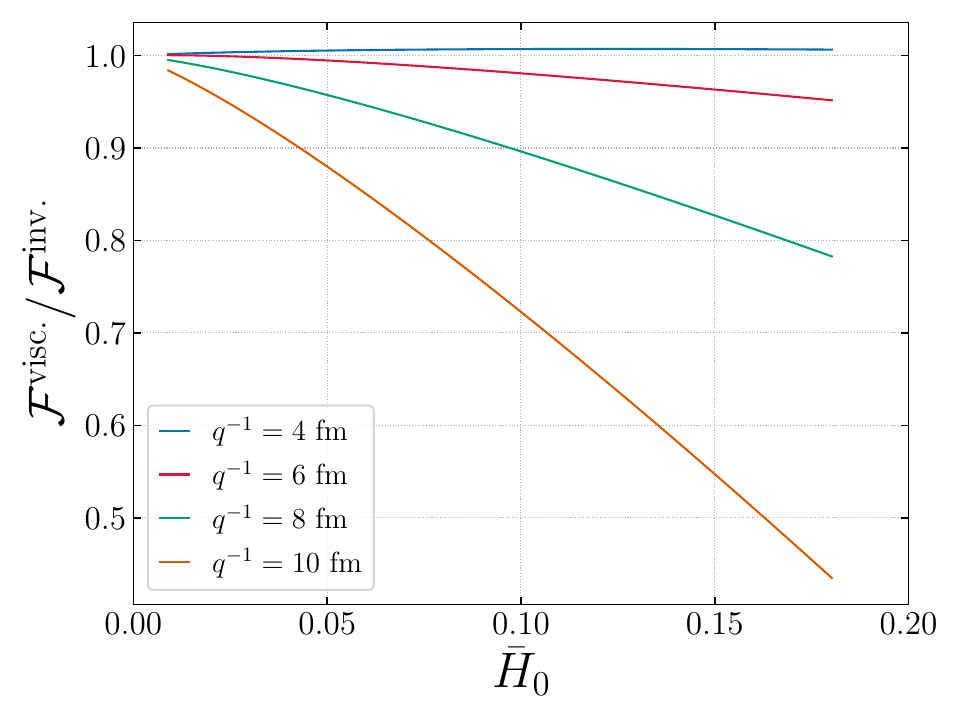}
    \includegraphics[width=0.45\linewidth]{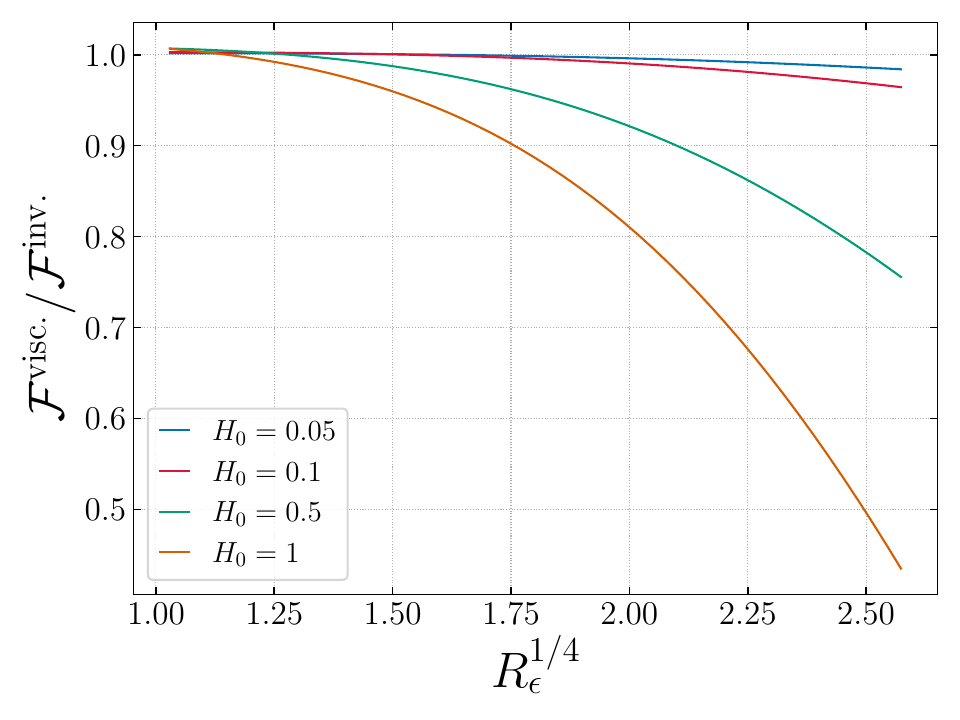}
    \caption{Ratio between $\cF$ for viscous and inviscid cases, computed numerically from Eq.~\eqref{eq:F_simp}, as a function of normalized viscosity $\bar H_0 = H_0/\hat T_0$ (left panel), for $q^{-1}\in(4,10)$ fm and of $R_\epsilon^{1/4} = \epsilon_h/\left(q\hat T_0\right)$ (right panel), for $H_0 \in (0.05, 1)$. We have set $\hat T_0 \simeq 5.5$ and $\epsilon_h \simeq 0.83$ GeV/fm$^3$.}
    \label{fig:Fbar_vs_invq_H0_viscous}
\end{figure}

In Fig.~\ref{fig:Fbar_vs_invq_inviscid}, we show $\cF$ for the inviscid case divided by its dimensionful pre-factor, as a function of $R_\epsilon^{1/4}$, which captures its full parametric dependence. We see that this quantity falls sharply with increasing $R_\epsilon^{1/4}$, i.e., it falls with increasing transverse system size $q^{-1}$, for fixed typical total energy at freeze-out, $\sim \epsilon_h/q^3$,  and fixed ratio $\epsilon_h/\hat T_0$. Turning to the viscous case, first note that the absolute value of the de Sitter temperature $|\hat T(g)| = \hat \epsilon(g)^{1/4}$ grows indefinitely for large $|g|$. For negative $g$ this is not an issue, since $x_\perp^2(\tau, g) > 0$ is not satisfied for sufficiently large negative $g$. For positive $g$, however, this condition is satisfied for arbitrarily large $g$ and the integrand goes as $g^{-1/2}$, making $\cF$ divergent. Not only this, but for sufficiently large $g$ one has $\hat T(g)<0$~\cite{Gubser:2010ui}. We thus impose a manual cutoff on the $g$ integration, setting it numerically to $g^\star$ defined by $\hat T(g^\star) = 0$.\footnote{For large $g$ this root behaves as $g^\star = \left(2(\hat T_0+H_0 a)/ H_0\right)^{3/2}$, where $a$ is a positive constant, i.e., $g^\star \to +\infty$ when $H_0 \to 0$ as expected.} In Fig.~\ref{fig:Fbar_vs_invq_H0_viscous}, we show $\cF$ in the viscous case divided by its inviscid counterpart, as a function of normalized shear viscosity strength $\bar H_0 = H_0/\hat T_0$, for fixed $q^{-1} \in (4,10)$ fm (left panel) and of $R_\epsilon^{1/4} = \epsilon_h^{1/4}/\left(q\hat T_0\right)$, for fixed $H_0 \in (0.05,1)$ (right panel). Starting with the left panel, for small enough transverse system size $q^{-1}$ we see a slight increase of energy flow for viscous flow in comparison with inviscid flow (e.g. blue line with $q^{-1} = 4$ fm), while for larger $q^{-1}$ this ratio is increasingly small. For $q^{-1} \gtrsim 5$ fm, the sensitivity to system size and viscosity becomes more pronounced, with the ratio dropping below $0.5$ for large enough $q^{-1}$ and viscosity strength. Looking at right panel, and taking into consideration the result for the inviscid case in Fig.~\ref{fig:Fbar_vs_invq_inviscid}, we conclude that increasing viscosity leads to a more rapid decrease in energy flow with transverse system size $q^{-1}$.

Let us now consider the viscous solution with azimuthal perturbations. Azimuthal perturbations to Gubser flow are discussed in detail in~\cite{Gubser:2010ze}, and for a large class of perturbations, a closed-form solution is not known and is instead given in terms of differential equations. For simplicity, we follow the approach of~\cite{Hatta:2014jva}, where a closed-form expression for azimuthal perturbations was derived in the $g \gg 1$ limit, corresponding to the early-time regime $q\tau \ll 1$. Let us define the perturbed energy density and velocity field as, respectively, $\epsilon = \epsilon_0(1 + \delta\epsilon)$ and $u^\mu = u_0^\mu + \delta u^\mu$, with the $0$ subscript indicating an unperturbed quantity. The perturbations then read~\cite{Hatta:2014jva}
\begin{align}\label{eq:azimuthal_perturbations}
     & \delta \epsilon = -4\left(1 + \frac{ H_0g^{2/3}}{2\hat T_0}\right)\epsilon_nA_n(x_\perp) \cos n\phi\,,\quad \quad \delta u^\tau = 0\,,
    \\
     & \delta u^\perp = \frac{3n(1-(qx_\perp)^2)}{4(qx_\perp)g}\epsilon_nA_n(x_\perp)\cos n\phi\,,\quad\quad \delta u^\phi = -\frac{3n\tau }{2x_\perp^2	}\epsilon_nA_n(x_\perp) \sin n\phi\,,
\end{align}
with $A_n(x_\perp) = \left(\frac{2qx_\perp}{1+(qx_\perp)^2}\right)^n$, and $\epsilon_n$ encodes the initial magnitude of the perturbation modes. The leading contributions to first order in $\epsilon_n$ are presented in appendix. There, we show that the perturbed result for Eq.~\eqref{eq:F_simp} can be written as 
\begin{align}
    \cF = \cF_0 + \epsilon_n \cF_n \cos n\phi
\end{align}
such that the relevant object to evaluate numerically is $\cF_n/\cF_0$, where $\cF_0$ is the unperturbed result. In fact, connecting this to the azimuthal EEC defined in Section~\ref{subsec:azimuthal_EEC}, as per Eqs.~\eqref{eq:EEC_harmonics} and~\eqref{eq:EEC_harmonics_ratio}, this ratio is roughly the ratio between the $n$th azimuthal EEC harmonic and the magnitude of the initial azimuthal deformation of the system $\epsilon_n$, i.e.,
\begin{align}
    \frac{1}{\epsilon_n}\frac{v_n^{\rm EEC}}{v_0^{\rm EEC}} = \frac{\cF_n}{\cF_0} + \cO(\chi^2)\,,
\end{align}
where $\chi$ is the angular separation between two detectors measuring the energy flow.

\begin{figure}[t!]
    \centering
    \includegraphics[width=0.45\linewidth]{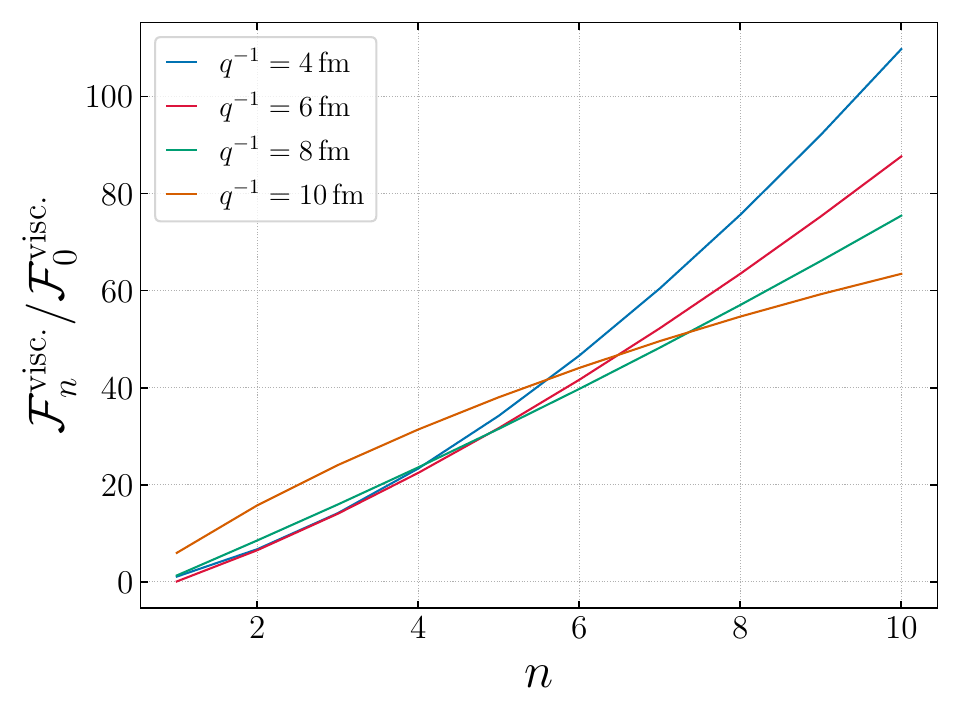}
    \includegraphics[width=0.45\linewidth]{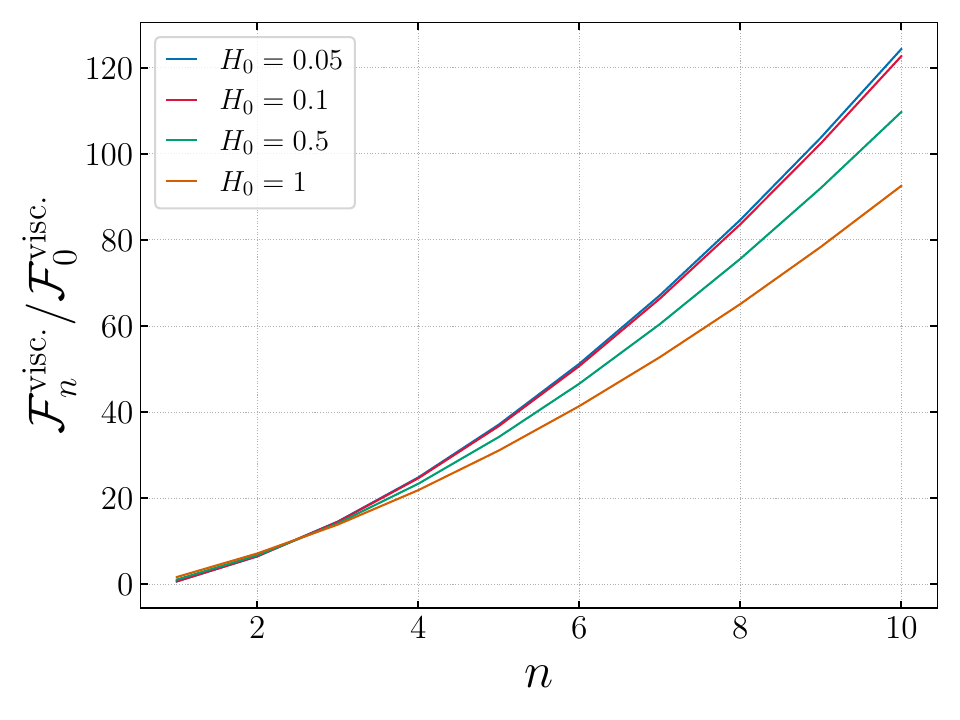}
    \caption{Ratio $ \cF_n/\cF_0$ for viscous Gubser flow as a function of harmonic number $n$, for multiple values of transverse size $q^{-1}\in(4,10)$ fm for fixed $H_0 = 0.5$ (left panel) and of viscosity $H_0\in(0.05,1)$ for fixed $q^{-1} = 4$ fm (right panel).}
    \label{fig:fn_over_f0_visc}
\end{figure}

In Fig.~\ref{fig:fn_over_f0_visc}, we show $\cF_n/\cF_0$ as a function of harmonic number $n$, for multiple values of $q^{-1}$ (left panel) for fixed $H_0 = 0.5$ and for multiple values of $H_0$ (right panel) for fixed $q^{-1} = 4$ fm. The first thing to notice is that the magnitude of the ratio becomes significant at large harmonic number $n$. To conclude any further, one would need a model to describe how $\epsilon_n$ scales with $n$. Then, looking at the left panel, for fixed viscosity $H_0$, we see that for lower harmonics the dependence on the transverse system size $q^{-1}$ is quite reduced, unless $q^{-1}$ is large enough (e.g. orange line). For such a large transverse size, the evolution with $n$ actually changes, becoming more mild than for the others. As for the viscosity dependence $H_0$ on the right panel, one also observes a very weak dependence for lower harmonics, with higher harmonics becoming more sensitive only for sufficiently large viscosity.

\section{Conclusions}\label{sec:conclusions}

In this work we initiated the study of energy-energy correlators measured on many-body QCD states. Focusing on the angular structure of the EEC in the collinear limit, we argued that the observable exhibits a sequence of regimes governed by distinct physical mechanisms, providing a unified picture of how microscopic and collective dynamics manifest themselves in detector-level observables.

At large angular separations the correlator is dominated by disconnected contributions associated with the collective hydrodynamic flow of the medium. In this regime the observable exhibits a universal scaling behavior that depends only on the geometric structure of the expanding fireball. We computed this contribution in the presence of Gubser flow, a simple analytical model for the hydrodynamic flow in HICs. Our results illustrate how the properties of the bulk medium are directly imprinted in the angular structure of the EEC, by considering the scaling with viscosity and transverse system size, and the role of perturbations on its otherwise azimuthally symmetric structure.

At smaller angular separations, connected contributions arising from fluctuations around local equilibrium become more relevant. Using general arguments based on the ETH and hydrodynamic correlation functions, we argued that this regime is governed by the hydrodynamic pole structure of the thermal propagators, leading to a distinct angular scaling behavior. As the angle is further reduced, the observable eventually transitions to the regime described by the light-ray OPE, where the correlator becomes sensitive to the microscopic dynamics of the underlying quantum field theory. This transition requires a detailed UV/IR matching procedure, which we leave for future work.

Together, these results suggest a characteristic angular sequence for EECs measured on bulk matter states, reflecting an IR/UV/IR structure associated with confinement dynamics, the light-ray OPE, and the hydrodynamic modes, respectively. Although the precise matching between these regimes remains an open problem, the parametric scaling behavior provides a robust framework for interpreting the observable.

These findings have direct implications for experimental studies of energy correlators in heavy-ion collisions. In particular, the large-angle scaling identified here should also appear in jet-based EEC measurements, where the observable receives contributions both from the perturbative jet cascade and from the collective response of the medium. As a consequence, the angular structure of the EEC provides a natural probe of the interplay between hard and soft physics in the quark–gluon plasma. An important avenue for future studies is therefore the systematic comparison of EEC measurements across collision systems of different sizes, ranging from $pp$ and $pA$ to $OO$, $AuAu$, and $PbPb$. Since the relative importance of collective dynamics increases with the size and lifetime of the produced medium, such a scan provides a natural way to disentangle contributions arising from the jet cascade and those associated with the bulk matter response.

\section*{Acknowledgments}
We thank G. Cuomo, E. Firat, A. Kurkela, S. Zhiboedov for useful discussions. The work of AVS is supported by Funda\symbol{"00E7}\symbol{"00E3}o para a Ci\symbol{"00EA}ncia e a Tecnologia (FCT) under contracts 
2023.15319.PEX (https://doi.org/10.54499/2023.15319.PEX) and 2022.06565.CEECIND, and by the Basque Government through grant IT1628-22. AVS would also like to acknowledge support from Ikerbasque, Basque Foundation for Science. 
The work of JMS is supported by FCT under contract PRT/BD/152262/2021 and project ERC-PT A-Projects ‘Unveiling’ (financed by PRR, NextGenerationEU), by European Research Council project ERC-2018-ADG-835105 YoctoLHC, by MCIN/AEI (10.13039/501100011033), by ERDF (grant PID2022-139466NB-C21), and by Consejería de Universidad, Investigación e Innovación, Gobierno de España and Unión Europea – NextGenerationEU under grant AST22 6.5. I.M. is supported by the Sloan Foundation.

\bibliographystyle{JHEP-2modlong.bst}

\bibliography{references.bib}

\appendix
\clearpage

\section*{Appendix}

\renewcommand{\theequation}{A\arabic{equation}}
\setcounter{equation}{0}

\subsection*{Explicit contributions for the azimuthally perturbed $\cF$}\label{sec:Appendix_Perturbations}
In this Appendix, we provide details on the calculation of the corrections due to azimuthal perturbations to the object $\cF$, as defined in Eq.~\eqref{eq:F_simp}. The azimuthal perturbations were stated in Eq.~\eqref{eq:azimuthal_perturbations} as derived in~\cite{Hatta:2014jva}. Although these perturbations are derived in the $g\gg 1$ limit (equivalently $\tau \ll 1$), we keep the remaining objects evaluated at finite $g$. In what follows, to ease the notation we define dimensionless time $t(g) = q\tau_h(g)$ and transverse coordinate $r(g) = (qx_\perp)(g) = \sqrt{2t(g)g + t(g)^2-1}$. The freeze-out condition then reads
\begin{align}
	t(g) = t_0(g)\left(1 + \frac{1}{4}\delta\epsilon(t(g),r(g),\phi)\right) \,,\qquad t_0(g) = R_\epsilon^{-1/4}(1+g^2)^{-1/3}(1+\bar H_0 g_v(g))\,,
\end{align}
with $R_\epsilon^{1/4} = \epsilon_h/\left(q\hat T_0\right)$ and $\bar H_0 = H_0 / \hat T_0$ as defined in the main text, $t_0(g)$ is the unperturbed freeze-out solution and $g_v(g)$ carries $g$ dependence of the the viscosity correction to $\hat T(g)$, i.e.
\begin{align}
    g_v(g) = g\left[(1+g^2)^{-1/6}-{}_2F_1\left(\frac{1}{2},\frac{1}{6};\frac{3}{2};-g^2\right)\right]\,.
\end{align}
Writing $\cF =  \cF_0 + \delta\cF$, where $\cF_0$ is the unperturbed result, the corrections that follow from Eq.~\eqref{eq:F_simp} read
\begin{align}
	\delta \cF = \frac{\epsilon_h}{q^3}\int_{-\infty}^{g^{\star}} dg\,\Big[\frac{t^3}{4}\Theta(r^2(g))\Theta( A_0^\tau(t,g))\frac{\delta_1 + \delta_2}{q\epsilon_h^2}\Big]_{t=t_0(g),\, r = r_0(g)}\,,
\end{align}
where $r_0(g) = \sqrt{2t_0(g)g + t_0(g)^2 - 1}$ and we have already imposed the cutoff $g^\star$ in the integration given by $\hat T(g^\star) = 0$, such that $\hat T(g) > 0 $ in the integration range as discussed in Section~\ref{sec:Gubser_flow_sec}. The corrections read
\begin{align}
	& \delta_1 = \frac{1}{4}\left(| A_{0,\perp}|\left(3\delta \epsilon + t(\partial_t \delta\epsilon)_g\right)+ t\delta\epsilon\, s_\perp \frac{dA_{0,\perp}}{dt}\right)\,,\nn
	& \delta_2 = s_\perp \left(-(r/q)\partial_\phi A^\phi+ \delta A_\perp\right)\,,
\end{align}
with $s_\perp = $ sign$(A_{0,\perp})$, $\frac{d}{dt}$ is to be understood as a total derivative, i.e., one accounts for the $t$ dependence in $r(t,g)$, and the unperturbed $ A_{0,\perp}$ is defined in Eq.~\eqref{eq:AGeneral_unperturbed}. The corrections in $\delta_1$ arise from evaluating the unperturbed integrand at a shifted freeze-out condition and the corrections in $\delta_2$ stem directly from the corrections to $ A^\mu$ from the azimuthal dependence of both velocity field and energy density. The first set of corrections reads
\begin{align}
	\delta_1 & = -2(q\epsilon_h^2)\left(1 + \frac{\bar H_0g^{2/3}}{2}\right)\epsilon_nA_n(r) \cos n\phi\Bigg[| A_{0,\perp}|\left(1 + \frac{t(g+t)}{2r^2(1+r^2)}\left(n+1 - r^2(n-1)\right)\right)\nn
	& +\frac{8r s_\perp}{3(1+g^2)(1+\bar H_0 g_v)}\left(1 + \bar H_0\left(g_v + \frac{g}{2}(1+g^2)^{-1/6}\right)\right)\Bigg]\,.
\end{align}
where $A_n(r) = \left(2r/(1+r^2)\right)^n$. For $\delta_2$, the correction to $A_\perp$ reads
\begin{align}
	\delta A_\perp &  \equiv (q\epsilon_h^2)\left(\delta A_{1,\perp} + 2\bar H_0 R_\epsilon^{-1/4}\delta A_{2,\perp}\right)\,,
\end{align}
where it is understood $\delta A_{2,\perp}$ is the term containing the dependence on the shear stress tensor $\sigma^{\mu\nu}$ and corrections to it. These two contributions explicitly read
\begin{align}
	& \delta A_{1,\perp} = \frac{1}{3tr} \epsilon_nA_n(r)\cos n\phi\Bigg[\frac{3n(1-r^2)}{g\sqrt{1+g^2}}\left(4(g+t) + \lambda(g)(1+g^2-r^2)\right)-\nn
	& -4\left(r^2\bar H_0 g^{-1/3} - \frac{4r^2 + (1+g^2)}{1+g^2}\left(1 + \frac{\bar H_0g^{2/3}}{2}\right)\frac{n(1-r^2)}{2g+t}\right)\Bigg]\,, \nn\nn
	& \delta A_{2,\perp} = \frac{n}{t^2r(1+g^2)}\epsilon_n A_n(r) \cos n\phi \,\Bigg[(1-r^2)(g+t)\Bigg( \lambda(g)\frac{(3r^2 + g(g+t))}{4g} + \nn
	& + \frac{4}{3}\left(1 + \frac{\bar H_0 g^{2/3}}{2}\right)\frac{g(g+t)}{(2g+t)\sqrt{1+g^2}}\Bigg) \nn
	& -\frac{1}{g^2t(1+g^2)}\left(\cP_0((1+g^2, t) + (g+t)\cP_1(1+g^2, t) + 2nt^2(g+t)(1+g^2)g^2\right)\Bigg]\,,
\end{align}
with 
\begin{align}
	\lambda(g) = \frac{\hat{\epsilon}'(g)}{\hat{ \epsilon}(g)} = -\frac{8g}{3(1+g^2)(1+\bar H_0 g_v)}\left[1 + \bar H_0\left(g_v - \frac{3(1+g^2)}{2g}g'_v\right)\right]\,,
\end{align}
and
\begin{align}
	& \cP_0(b,t)=2 b^3 t+13 b^2 t^3-12 b^2 t-12 b t^5-22 b t^3+52 b t+26 t^5-10 t^3-36 t\,,\nn
	& \cP_1(b,t)=5 b^2 t^2-2 b^2+16 b t^4-66 b t^2+4 b-31 t^4+83 t^2-10\,.
\end{align}
Finally, the correction involving $A^\phi$ reads 
\begin{align}
	& -(r/q)\partial_\phi A^\phi = 
	q\epsilon_h^2\Bigg[\frac{1}{3}\left(\frac{3}{\sqrt{1+g^2}}\left(4(g+t) + \lambda(g)(1+g^2)\right) + 2\left(1 + \frac{\bar H_0 g^{2/3}}{2}\right)\right)\nn
	& -\frac{2\bar H_0 R_\epsilon^{-1/4}}{t\sqrt{1+g^2}}\left(\frac{3C_\phi}{2\sqrt{1+g^2}}- \frac{2g}{3}\left(1 + \frac{\bar H_0 g^{2/3}}{2}\right)\right)\Bigg]\frac{2n^2}{r}\epsilon_n A_n(r) \cos n\phi\,,
\end{align}
where
\begin{align}
	C_\phi = -(1+t^2) + n\frac{(1-r^2)(g+t)(4g+t)}{2g(2g+t)}\,.
\end{align}
Thus, one concludes $\delta \cF =  \epsilon_n\cF_n\cos n\phi$, with the $n$ dependence in the integrand inside $\cF_n$ being given by the product of $A_n(r)$ and a third degree polynomial in $n$.

\end{document}